\documentclass[11pt]{article}
\usepackage{listings}
\usepackage{arxiv}
\usepackage[numbers]{natbib}
\usepackage{algorithm}
\usepackage{algpseudocode}
\usepackage{times}
\usepackage{latexsym}
\usepackage{booktabs}
\usepackage{amsmath}
\usepackage{amssymb}
\usepackage[table]{xcolor}
\usepackage{listings}
\usepackage{hyperref}

\usepackage[T1]{fontenc}

\usepackage[utf8]{inputenc}

\usepackage{microtype}

\usepackage{graphicx}
\usepackage[most]{tcolorbox}
\usepackage{fancyvrb}
\usepackage{fvextra}
\newtcolorbox{promptbox}[1]{
  enhanced, breakable,
  colback=gray!3, colframe=black!72, coltitle=white,
  fonttitle=\bfseries\small, title={#1},
  boxrule=0.6pt, arc=2pt, left=5pt, right=5pt, top=3pt, bottom=3pt,
}
\DefineVerbatimEnvironment{promptverb}{Verbatim}{breaklines=true,breakanywhere=true,fontsize=\footnotesize}

\title{Constrained Dominant Sets for Multimodal Document Question Answering}

\author{%
  Ambuj Mehrish \\
  Ca' Foscari University of Venice \\
  \texttt{ambuj.mehrish@unive.it} \\
  \And
  Sebastiano Vascon \\
  Ca' Foscari University of Venice \\
  \texttt{sebastiano.vascon@unive.it} \\
}

\begin{document}
\maketitle
\begin{abstract}
Long multimodal document question answering is limited by which evidence reaches the reader, rather than by the quantity retrieved. In lengthy documents, findings often recur across figures, captions, and introductory sentences, causing similarity-based retrievers in modern multimodal retrieval-augmented generation (RAG) systems to allocate resources to near-duplicates while overlooking complementary evidence. This work introduces a retriever that selects evidence as a Constrained Dominant Set (CDS) on a query-augmented affinity graph, offering three advantages that similarity ranking does not. First, the query is encoded as a hard structural constraint, ensuring that every selected element is directly connected to the question through the cluster anchor. Second, the relevance–redundancy balance is determined automatically by a spectral bound, eliminating the need for manually tuned trade-offs required by diversity-aware selectors. Third, the selection process achieves a global equilibrium via replicator dynamics, thereby avoiding the distortions introduced by greedy heuristics. The method is inherently graph-based and does not require training. Using a Qwen3-VL-32B reader, CDS establishes a new state of the art on VisDoMBench ($66.99$ average) and improves over the no-retrieval baseline by $37.1$ points on VisDoMBench and $4.8$ on MMLongBench-Doc.
\end{abstract}

\section{Introduction}
\label{sec:intro}

Vision–language models (VLM) encounter persistent challenges in processing long multimodal documents. On MMLongBench-Doc~\citep{ma2024mmlongbench}, which consists of 135 PDFs with an average length of 47.5 pages, GPT-4o achieves an F1 score of only 44.9. \citet{liu2024lost} identify a U-shaped attention curve, indicating that evidence situated in the middle of a long context is systematically under-utilized. This phenomenon can result in inferior multi-document question-answering performance compared to the closed-book baseline. Therefore, the principal limitation lies not in the quantity of accessible information, but in the selection process.

Most current multimodal RAG systems provide a flat top-$B$ list ranked by query–candidate similarity~\citep{faysse2024colpali,suri2025visdom,wang2025vidorag,wan2025mmgraphrag,guo2024lightrag,guo2025rag,nguyen2025ma,du2026g}. For queries related to a chart or table in a lengthy report, these retrievers frequently return the visual element, its caption, the introductory sentence, and the paragraph reiterating the finding. This method allocates multiple retrieval slots to a single fact, thereby limiting the inclusion of methodology, related results, or qualifying footnotes. Such redundancy is by design: long multimodal documents are structured to restate findings across modalities to support diverse reading patterns, and similarity-based retrieval further amplifies this engineered redundancy.

The information-retrieval community has long acknowledged that relevance alone is insufficient. Maximal Marginal Relevance (MMR)~\citep{carbonell1998use} incorporates a redundancy penalty into the relevance score, while Determinantal Point Processes (DPPs)~\citep{kulesza2012determinantal,chen2018fast} favor subsets with greater feature diversity. However, both approaches are heuristic in the budget-constrained maximum a posteriori (MAP) regime: MMR employs a hand-tuned linear scalarization $\lambda\,\mathrm{sim}(q, d) - (1-\lambda)\max_{j\in S}\mathrm{sim}(d, d_j)$ optimized greedily, causing the initial selection to constrain subsequent choices. DPP MAP is NP-hard, and standard pipelines revert to the same greedy approximation~\citep{gillenwater2012near}. Importantly, neither method provides a native mechanism to enforce coherence between the selected set and the query. The query serves as a scoring signal rather than a constraint. While this distinction may be negligible for flat candidate lists, it becomes significant for graph-structured candidate sets in long multimodal documents, where captions, figures, and passages often form near-identical clusters.

We reformulate evidence selection as a \emph{constrained dominant-set} (CDS) problem~\citep{bulo2017dominant} on a query-augmented affinity graph. Dominant sets~\citep{pavan2007dominant,bulo2017dominant} generalize maximal cliques by representing strict local maxima of a quadratic program on the simplex, while constrained dominant sets~\citep{zemene2016interactive,zemene2018dominant} additionally require the cluster to include a user-specified seed. In this framework, the query and any pinned memory notes constitute the constraint set $S$. Query--note edges are encoded as similarity, and note--note edges as dissimilarity. The resulting program is solved using replicator dynamics from evolutionary game theory~\citep{pelillo1998replicator,bomze1997evolution}. The fixed point yields a soft membership vector whose support defines a cluster anchored on the query, characterized by high mutual relevance and low mutual redundancy. This approach requires no hyperparameter tuning, kernel design, or greedy locking; the selection emerges as the dynamics' equilibrium.

CDS methods have demonstrated strong empirical performance in computer vision, including image segmentation~\citep{zemene2018dominant}, image retrieval~\citep{ALEMU2020103862}, person re-identification~\citep{alemu2019deep}, multi-target tracking~\citep{tesfaye2019multi}, and geo-localization~\citep{zemene2018large}. However, these methods have not yet been applied to natural language processing (NLP) retrieval tasks. Long multimodal RAG presents a suitable context for this approach: candidate sets are sufficiently small for rapid convergence of the dynamics, the link structure is complex enough to benefit from graph-native methods, and the evidence budget is sufficiently constrained ($B=10$, consistent with modern multimodal RAG systems~\citep{du2026g,wang2025vidorag,suri2025visdom,cho2024m3docrag}) to make the relevance and diversity trade-off critical. The impact of varying $B$ is analyzed in Appendix~\ref{app:evibud}.

\paragraph{Contributions.}
\begin{itemize}
\item A \textbf{query-as-constraint} formulation is introduced, in which the query is incorporated into the constraint diagonal of the CDS objective. This approach enables query coherence to emerge as an equilibrium property rather than as a result of tuning trade-offs. The formulation generalizes to multi-query and pinned-memory settings without modification.
\item The \textbf{first application of constrained dominant sets to multimodal RAG} is demonstrated, instantiated on an agentic-memory-style note graph~\citep{xu2026mem} that incorporates note--note dissimilarity edges, query--note similarity edges, and employs replicator dynamics as the solver.
\item \textbf{State-of-the-art (SOTA) results on VisDoMBench~\cite{suri2025visdom} and competitive performance on MMLongBench-Doc~\citep{ma2024mmlongbench}.} Using a Qwen3-VL-32B reader, the method attains $66.99$ average accuracy on VisDoMBench, outperforming $G^2$-Reader, ViDoRAG, MMGraphRAG, LightRAG, RAG-Anything, and a standalone GPT-5. 
\end{itemize}

\section{Related Work}
\label{sec:related}
\paragraph{Multimodal RAG.} Recent research extends RAG~\citep{lewis2020retrieval} to visually rich documents using several complementary strategies. These include page-image retrieval with late-interaction matching (ColPali~\citep{faysse2024colpali}, M3DocRAG~\citep{cho2024m3docrag}) and single-vector visual-document encoding (DSE~\citep{ma2024unifying}), text-visual fusion with consistency-constrained or iterative reasoning (VisDoMRAG~\citep{suri2025visdom}, ViDoRAG~\citep{wang2025vidorag}), entity- or knowledge-graph traversal (GraphRAG~\citep{edge2024local}, LightRAG~\citep{guo2024lightrag}, MMGraphRAG~\citep{wan2025mmgraphrag}, RAG-Anything~\citep{guo2025rag}), multi-agent chain-of-thought (MA-RAG~\citep{nguyen2025ma}), dual evolving graphs ($G^2$-Reader~\citep{du2026g}, previously the state of the art on VisDoMBench), and retrieval-aware tuning for long documents (M-LongDoc~\citep{chia2025m}). In most approaches, evidence is selected by ranking candidates against the query and selecting the top-$B$ using cosine or learned similarity. Even graph-based systems primarily use the graph to produce or enrich candidates, rather than to select a budget-limited subset, with the final selection still based on a flat top-$B$ ranking. The proposed method adopts this candidate-production pipeline but replaces the final selection step with constrained graph clustering.

\paragraph{Long-document and multimodal QA benchmarks.} Document question answering (QA) has evolved from single-page extraction~\citep{mathew2021docvqa} to multi-page reasoning~\citep{tito2023hierarchical} and, more recently, to long-context multimodal benchmarks that address distinct modalities. These modalities include slides (SlideVQA~\citep{tanaka2023slidevqa}), scientific figures and tables (SPIQA~\citep{pramanick2024spiqa}, PaperTab~\citep{hui2024uda} and FetaTab~\citep{nan2022fetaqa}), charts (SciGraphQA~\citep{li2023scigraphqa}), and multi-document settings (VisDoMBench~\citep{suri2025visdom}). MMLongBench-Doc~\citep{ma2024mmlongbench} emphasizes single-document long-context comprehension, while M-LongDoc~\citep{chia2025m} extends the average document length beyond 200 pages, and LongDocURL~\citep{deng2025longdocurl} integrates understanding, reasoning, and locating over multimodal long documents.

\paragraph{Memory-augmented LLM systems.} The candidate set is constructed following the approach of agentic memory systems, such as MemGPT~\citep{packer2023memgpt}, MemoryBank~\citep{zhong2024memorybank}, and A-MEM~\citep{xu2026mem}, where each document element is transformed into a memory note enriched with summaries, keywords, tags, and explicit links. These links provide the affinity signal that drives the CDS dynamics. While previous work has focused on constructing and evolving the memory store, this work addresses the complementary problem of selecting a query-conditioned, budget-respecting, and mutually coherent subset from it.

\section{Methodology}
\label{sec:methodology}
We propose a document-grounded multimodal question answering pipeline for long documents containing text, figures, tables, and other visually
grounded content. The method represents each document as a multimodal graph, retrieves a compact and coherent evidence subgraph using CDS optimization, and generates the final answer with a VLM conditioned only on the retrieved evidence.
\begin{figure*}
\centering
\includegraphics[width=\textwidth]{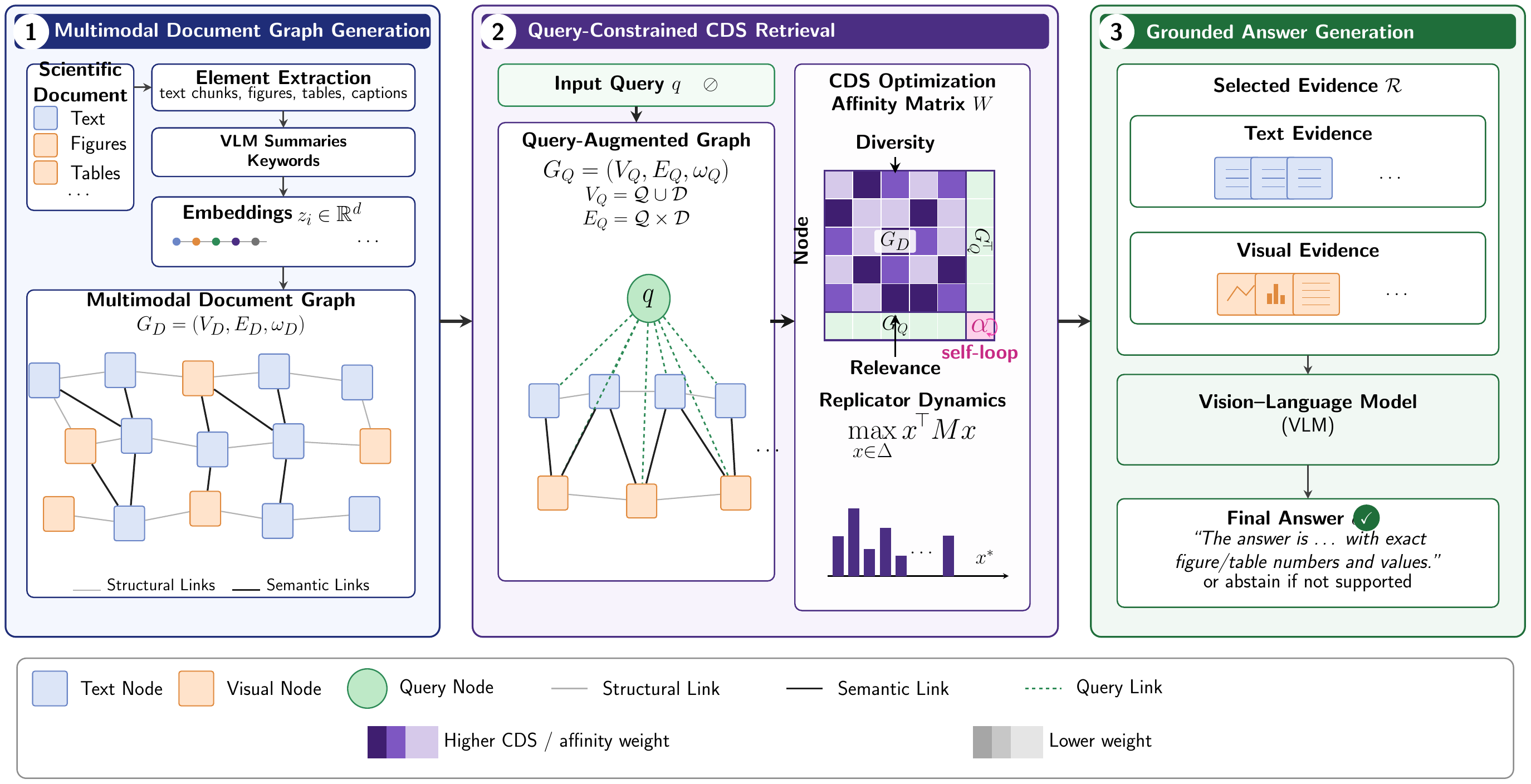}
\caption{Overview of the proposed pipeline. \textbf{(1)} The document is parsed into atomic nodes (text, figures, tables), each embedded as $\mathbf{z}_i \in \mathbb{R}^d$ and connected into a multimodal graph $G_D$ via structural and VLM-verified semantic links. \textbf{(2)} The query $q$ is added as a constraint vertex with a spectral self-loop $\alpha$, forming the query-augmented affinity matrix $M$ that combines note--note dissimilarity (diversity) with query--note similarity (relevance). Replicator dynamics solve $\max_{\mathbf{x}\in\Delta}\mathbf{x}^\top M \mathbf{x}$, yielding a soft membership $\mathbf{x}^\star$ anchored on $q$. \textbf{(3)} The top-$B$ nodes by $x_i^\star$ are passed to a VLM reader that generates the final answer grounded strictly in the retrieved evidence.}
\label{fig:pipeline}
\end{figure*}
\subsection{Problem Formulation}
\label{sec:problem_formulation}

Let $D=\{e_1,\dots,e_N\}$ be a long multimodal document decomposed into $N$ elements, where each $e_i$ is a self-contained unit of evidence such as  text passage, figure, table or caption. Given a question $q$, the objective is to produce an answer $\hat a$ that is grounded in a subset of these elements. In this work, we adopt a \emph{retrieve-then-read} formulation: a retriever selects an evidence set $R\subseteq D$, and a reader VLM generates an answer based on the question and the selected evidence, $\hat a = f_{\mathrm{VLM}}(q, R)$.

The primary challenge is selecting $R$. Relevance alone is insufficient because the most relevant elements are often near-duplicates, such as a passage and its caption or repeated content, which results in redundant information. Effective evidence must be both \emph{relevant} to the question and mutually \emph{non-redundant/diverse}, ensuring that the selected elements collectively cover complementary information. Retrieval is therefore formalized as follows:
\begin{equation}
R^{\star} = \arg\max_{R\subseteq D} \;\mathrm{Rel}(R,q)\;+\;\mathrm{Div}(R),
\label{eq:problem_formulation}
\end{equation}

$\mathrm{Rel}(R,q)$ quantifies query relevance, $\mathrm{Div}(R)$ rewards diversity by minimizing mutual redundancy among selected elements. Both $\mathrm{Rel}(R,q)$ and $\mathrm{Div}(R)$ are quadratic terms as they relate sets, the first considers the relations between $\mathcal{D}$ and $\mathcal{Q}$ while the latter considers $\mathcal{D}$ with $\mathcal{D}$. Equation~\eqref{eq:problem_formulation} looks for a subset $R\subseteq D$ and is formulated as a constrained weighted maximal clique extraction problem (CDS) on a query-augmented graph constructed from the document elements. In this graph, edge weights represent query--element relevance and element--element dissimilarity, with the query incorporated as a constraint vertex to anchor the selected cluster. The reader receives only $q$ and $R$ and is instructed to answer strictly on the basis of the provided evidence, to preserve exact figure or table identifiers and numerical values, and to abstain when support is lacking.

\subsection{CDS Graph Construction}
\label{sec:cds-graph}
We use CDS as the retrieval operator over a query-augmented multimodal graph. The task is to extract a (possibly tight) partition of the nodes of a graph that contains a designated node: in our setting, the graph's nodes are the document's elements and the designated node is the query. CDS model addresses exactly this task by extracting a subset of nodes that contains a given seed; it requires a similarity graph and a set of query nodes, and returns the partition. We therefore work with two sets: the document set $\mathcal{D}$ and the query set $\mathcal{Q}=\{q\}$.

\paragraph{Document--Document graph.} For each document, a multimodal graph $G_D=(V_D,E_D,\omega_D)$ is constructed over its elements. The source document is processed by a multimodal parser (e.g., MinerU~\citep{wang2024mineru} or DeepSeek-OCR~\citep{wei2025deepseek}) that extracts text passages and visual elements while preserving layout and reading order. These elements are segmented into atomic units, forming the node set $V_D=\mathcal{D}$. Each node $v_i\in V_D$ stores its raw content $c_i$ (a text passage or a figure, table, or image crop with its caption and surrounding context), a VLM-generated summary $s_i$, and a keyword set $k_i$. To integrate heterogeneous text and visual elements into a shared semantic space, each node is represented by a textual surrogate $m_i = s_i \,\Vert\, \text{``keywords:''} \,\Vert\, k_i$, where $\Vert$ denotes concatenation, and is encoded as $\mathbf{z}_i = f_{\mathrm{emb}}(m_i)\in\mathbb{R}^{d}$. Edge candidates $E_D \subseteq V_D \times V_D$ are initially proposed based on document structure (such as adjacent text chunks, caption--figure and caption--table pairs, explicit ``Figure'' or ``Table'' references, and page proximity) and are retained only if a VLM verifies them as genuine relations (including direct reference, elaboration, part--whole, causal, or contextual dependency). Following $G^2$-Reader~\citep{du2026g}, this construction is iterated to progressively refine the graph, with three evolution iterations employed. Edge weights between node $i$ and $j$ is defined as $\omega_D(i,j)=1-\cos(\mathbf{z}_i,\mathbf{z}_j)$\footnote{The note block can be optionally scaled by a diversification strength $\beta$, where $\beta = 1$ restores the default. This parameter is systematically varied in Appendix~\ref{app:evibud} (Table~\ref{tab:beta})}. Two properties are notable: (i) this is a dissimilarity graph, where more diverse nodes are more strongly connected, and (ii) it is static, computed once and reused for all queries. The graph is stored as a pair-wise dissimilarity matrix $W \in \mathbb{R}^{|\mathcal{D}|\times|\mathcal{D}|}$. 
The graph $G_D$ captures the term $\mathrm{Div}(R)$.
\paragraph{Query--Document graph.} Given the query set $\mathcal{Q}$ and the document set $\mathcal{D}$, we construct a bipartite graph $G_Q=(V_Q,E_Q,\omega_Q)$ where $V_Q=\mathcal{Q}\cup\mathcal{D}$, $E_Q=\mathcal{Q}\times\mathcal{D}$, and
\[
\omega_Q(i,j)=
\begin{cases}
0, & i,j\in \mathcal{Q},\\
\cos(\mathbf{z}_i,\mathbf{z}_j), & \text{otherwise},
\end{cases}
\]
with the query encoded using the same encoder, $\mathbf{z}_q = f_{\mathrm{emb}}(q)$. This is a \emph{similarity} graph: the more similar a query $q\in\mathcal{Q}$ is to a document node $d\in\mathcal{D}$, the more strongly they are connected. The graph is stored in a similarity matrix $S\in \mathbb{R}^{|\mathcal{Q}|\times|\mathcal{D}|}$; since $\mathcal{Q}=\{q\}$, this reduces to a row vector $S\in\mathbb{R}^{1\times|\mathcal{D}|}$. A sparsification threshold $\tau$ may be applied to prune weak query--node edges, specifically those for which $\cos(\mathbf{z}_q, \mathbf{z}_i) < \tau$. The default setting is $\tau = 0$, and its effect is analyzed in Appendix~\ref{app:tau}. The graph $G_Q$ captures the quadratic term $\mathrm{Rel}(R,q)$.

The next step is to combine the two graphs such that the CDS can extract the relevant nodes from $\mathcal{D}$ given $\mathcal{Q}$. Remember that $G_D$ captures the most diverse nodes, while $G_Q$ models the most relevant nodes to the query. We combine the two graphs $G_D$ and $G_Q$ into $G=(V,E,\omega)$, where $V=V_D \cup V_Q$, $E=E_D \cup E_Q$ and $\omega$ is defined as:
\begin{equation}
\setlength{\arraycolsep}{2pt}
\small
\omega_{i,j} = \begin{cases}
\omega_D(i,j) = 1 - \cos(\mathbf{z}_i,\mathbf{z}_j), & i \in \mathcal{D} \text{ and } j \in \mathcal{D}, \\
0,                              & i = j \text{ and } i \in \mathcal{D}, \\
\alpha,                         & i = j \text{ and } i \in \mathcal{Q}, \\
\omega_Q(i,j) = \cos(\mathbf{z}_i,\mathbf{z}_j),      & \text{otherwise}.
\end{cases}
\end{equation}
where $\alpha = \lambda_{\max}(W) + \varepsilon$, $\lambda_{\max}(\cdot)$ is the largest eigenvalue of the matrix $W$, and \(\varepsilon>0\) is a small margin. We use $\varepsilon=10^{-3}$. The CDS extracts sequentially the maximal cliques from the graph $G$ until all the query nodes in $\mathcal{Q}$ get partitioned. The value $\alpha$ on the self-loop of the node $q$ ensures, by construction, that $q$ will be extracted as part of a maximal clique. The graph $G$ is then represented as a pairwise matrix $M \in \mathbb{R}^{|V|\times|V|}$. The matrix $M$ captures both the $\mathrm{Div}(R)$ and the $\mathrm{Rel}(R,q)$ (see figure \ref{fig:pipeline} for a graphical interpretation).

To extract the subset $R$ that is both diverse and relevant in $G$, we optimize the following quadratic assignment problem using the replicator dynamics (see \citep{zemene2018dominant} for proofs and details):
\begin{equation}
\mathbf{x}^{\star} = \arg\max_{\mathbf{x}\in\Delta} \mathbf{x}^{\top}M\mathbf{x}.
\label{eq:cds_objective}
\end{equation}
where $\Delta$ is the standard $|V|-$dimensional simplex (i.e. $\sum_i x_i = 1$). The solution \(\mathbf{x}^{\star}\) assigns a soft membership weight to each document and query vertex to the maximal clique. Its support is $\sigma(\mathbf{x}^{\star}) = \{i:x_i^{\star}>0\}$. At convergence, the query vertex is removed from the support, and document nodes are ranked by  $\mathrm{score}(v_i)=x_i^{\star}$.

\paragraph{Selection and generation.} Solving Eq.~\eqref{eq:cds_objective} via replicator dynamics yields a soft membership weight $\mathbf{x}^{\star}$ to each vertex. We discard the query vertex and rank documents vertex by $\mathrm{score}(v_i)=x_i^{\star}$, forming the evidence set from the top-$B$:
\begin{equation*}
R = \operatorname{Top\text{-}}B_{\,v_i\in\mathcal{V}}\big(\mathrm{score}(v_i)\big).
\end{equation*}
The VLM then receives \emph{only} the question $q$ and $R$: for text nodes, the passage (truncated if long); for visual nodes, the image with its caption or generated summary. The answer is thus grounded in the retrieved multimodal evidence rather than in auxiliary reasoning chains.

\section{Dataset and Baselines}
\subsection{Datasets}
\label{sec:datasets}
We conduct evaluations on two multimodal long-document QA benchmarks, adhering to the VisDoM~\citep{suri2025visdom} and $G^2$-Reader~\citep{du2026g} protocols to ensure direct comparability with prior state-of-the-art methods. \textbf{VisDoMBench} comprises a heterogeneous suite that encompasses the principal modalities of real-world long-document QA: \textbf{FetaTab} and \textbf{PaperTab} (table-grounded QA over web and scientific tables), \textbf{SciGraphQA} (chart-grounded QA over scientific graphs), \textbf{SlideVQA} (slide decks containing mixed text, figures, and tables), and \textbf{SPIQA} (figure- and table-grounded QA over scientific papers). Each query is associated with a candidate set limited to five documents (one gold and up to four distractors), ensuring that retrieval occurs within a realistic multi-document context. This setting intensifies the redundancy challenge, as a single fact may appear in a passage, its caption, and a downstream summary across the five parsed documents.

We also report results on \textbf{MMLongBench-Doc}~\citep{ma2024mmlongbench}, where each query is grounded in a single, very long document. This benchmark is evaluated separately because the primary retrieval challenge is within-document redundancy, such as repeated passages, caption and passage duplicates, and recurring boilerplate, rather than cross-document overlap. This setting provides a rigorous test of CDS performance under an extended per-query note set.

For both benchmarks, every source PDF is parsed offline with MinerU~\citep{wang2024mineru} into a structured stream of text chunks (recursive splitting, chunk size $3000$, overlap $300$) and visual elements (figures, tables, charts) with captions and a $\pm 1000$-word local context window. Each element becomes a memory note enriched by a VLM with a content summary, keywords, tags, and links to related notes (Section~\ref{sec:cds-graph}). The per-query note set is the union of notes from all candidate documents and forms the vertex set of the affinity graph that every retrieval method below operates on.
\begin{table*}[t]
\centering
\scriptsize
\setlength{\tabcolsep}{5pt}
\renewcommand{\arraystretch}{1.10}
\caption{Main results on the full VisDoMBench benchmark. All results are averaged over three runs, with ``$\pm$'' denoting standard deviation. $\dagger$ indicates systems using OpenAI \texttt{text-embedding-3-small} as the retrieval backbone; $\ddagger$ indicates our method using the open \texttt{nomic-embed-text-v1.5} encoder. \textbf{Bold} marks the best per-column score.}
\resizebox{\textwidth}{!}{%
\begin{tabular}{@{}lcccccc@{}}
\toprule
\textbf{Model} & \textbf{SPIQA} & \textbf{FetaTab} & \textbf{PaperTab} & \textbf{SciGraphQA} & \textbf{SlideVQA} & \textbf{Average} \\
\midrule
\textbf{GPT-5} & 55.22 $\pm$ 0.09 & 63.94 $\pm$ 0.31 & 37.08 $\pm$ 0.12 & 64.08 $\pm$ 0.32 & 45.06 $\pm$ 0.10 & 53.08 $\pm$ 0.10 \\
\textbf{Qwen3-VL-32B}$^{\dagger}$ & 29.86 $\pm$ 0.08 & 37.39 $\pm$ 0.36 & 34.32 $\pm$ 0.27 & 23.06 $\pm$ 0.22 & 24.87 $\pm$ 0.24 & 29.90 $\pm$ 0.11 \\
\textbf{Deepseek-OCR} & 63.60 $\pm$ 0.40 & 70.32 $\pm$ 0.12 & 51.58 $\pm$ 0.24 & 61.91 $\pm$ 0.40 & 65.69 $\pm$ 0.12 & 62.62 $\pm$ 0.13 \\
\textbf{RAGAnything} & 67.69 $\pm$ 0.96 & 57.76 $\pm$ 0.24 & 42.02 $\pm$ 1.35 & 41.60 $\pm$ 2.60 & 52.18 $\pm$ 0.49 & 52.25 $\pm$ 0.63 \\
\textbf{MA-RAG} & 45.52 $\pm$ 0.22 & 27.70 $\pm$ 0.19 & 33.43 $\pm$ 0.45 & 29.32 $\pm$ 0.25 & 29.40 $\pm$ 0.21 & 33.07 $\pm$ 0.13 \\

\textbf{GraphRAG} & 62.65 $\pm$ 0.20 & 61.35 $\pm$ 0.19 & 42.90 $\pm$ 0.00 & 65.76 $\pm$ 0.38 & 21.68 $\pm$ 0.00 & 50.87 $\pm$ 0.09 \\
\textbf{LightRAG} & 73.88 $\pm$ 0.00 & 64.71 $\pm$ 0.38 & 51.02 $\pm$ 0.04 & \textbf{75.00 $\pm$ 0.01} & 29.63 $\pm$ 0.01 & 58.85 $\pm$ 0.08 \\
\textbf{MMGraphRAG} & 69.91 $\pm$ 0.23 & \textbf{72.40 $\pm$ 0.55} & 56.36 $\pm$ 0.58 & \underline{64.11 $\pm$ 0.25} & 54.20 $\pm$ 0.15 & 63.40 $\pm$ 0.18 \\
\textbf{VisDoMRAG} & 75.44 $\pm$ 0.00 & 61.02 $\pm$ 0.50 & 56.21 $\pm$ 0.15 & 63.36 $\pm$ 0.14 & 69.03 $\pm$ 0.36 & 65.01 $\pm$ 0.13 \\
\textbf{ViDoRAG} & 68.18 $\pm$ 0.46 & 58.74 $\pm$ 0.38 & 43.67 $\pm$ 0.15 & 37.86 $\pm$ 0.14 & \underline{71.71 $\pm$ 0.11} & 56.03 $\pm$ 0.13 \\
\textbf{$G^2$-Reader}$^{\dagger}$ & 73.19 $\pm$ 0.21 & 66.89 $\pm$ 0.11 & \underline{57.10 $\pm$ 0.21} & 61.56 $\pm$ 0.11 & \textbf{72.31 $\pm$ 0.00} & \underline{66.21 $\pm$ 0.07} \\
\textbf{Qwen3-VL-32B + CDS}$^{\ddagger}$ & \textbf{78.85 $\pm$ 0.00} & \underline{70.90 $\pm$ 0.00} & \textbf{65.59 $\pm$ 0.01} & 57.45 $\pm$ 0.31 & 62.17 $\pm$ 0.38 & \textbf{66.99 $\pm$ 0.10} \\

\bottomrule
\end{tabular}%
}
\label{tab:main_results}
\end{table*}
\begin{table}[t]
\centering
\footnotesize
\caption{Results on MMLongBench-Doc. Accuracy (\%); ``+ CDS'' is our retriever,
reported as mean $\pm$ std over 3 runs. The plain rows are the Single-VLM
(no-retrieval) baseline.}
\label{tab:mmlb}
\resizebox{0.5\textwidth}{!}{%
\begin{tabular}{llc}
\toprule
Model & Params & Accuracy \\
\midrule
Qwen3-VL-32B        & 32B & 40.19 \\
Qwen3-VL-32B + CDS  & 32B & \textbf{45.01 $\pm$ 0.39} \\
\addlinespace
Qwen2.5-VL-7B       & 7B  & 28.36 \\
Qwen2.5-VL-7B + CDS & 7B  & \textbf{32.30 $\pm$ 0.10} \\
\addlinespace
GLM-4.1V            & 9B  & \textbf{41.04} \\
GLM-4.1V + CDS      & 9B  & 39.15 $\pm$ 0.20 \\
\bottomrule
\end{tabular}}
\end{table}

\subsection{Baselines}
\label{sec:baselines}
To isolate the contribution of CDS-based selection, our \emph{selection-rule} Baselines and the no-retrieval control are executed using an identical pipeline, ensuring consistency across experimental conditions. same parser (MinerU), note-construction and graph-evolution, text encoder (\texttt{nomic-embed-text-v1.5}, 768-dim, $L_2$-normalized)~\citep{nussbaum2024nomic}, evidence budget ($B{=}10$), reader VLM, and greedy decoding ($ \tau{=}0$), ensuring that the only variable is the selection mechanism. Published systems are not re-implemented; instead, their scores are cited from $G^2$-Reader~\citep{du2026g}. These systems are evaluated under their original configurations, which utilize a more advanced retrieval backbone\footnote{OpenAI text-embedding-3-small} but the same candidate-set protocol, evidence budget, and LLM-as-judge pipeline adopted here. 
\begin{table}[t]
\centering
\setlength{\tabcolsep}{4pt}
\caption{Cosine vs.\ CDS on the subset (50 queries/dataset), the same 250 samples used in $G^2$-Reader's ablation. Accuracy (\%). $G^2$-Reader reports $68.0$ on this subset~\citep{du2026g}.}
\label{tab:sample250}
\resizebox{0.6\textwidth}{!}{%
\begin{tabular}{lcccccc}
\toprule
Method & SPIQA & FetaTab & PaperTab & SciGraphQA & SlideVQA & Avg. \\
\midrule
Cosine (top-$B$) & 88.0 & 73.5 & 66.0 & 48.0 & 52.0 & 65.5 \\
\textbf{CDS} & \textbf{94.0} & \textbf{77.6} & 66.0 & \textbf{50.0} & \textbf{60.0} & \textbf{69.5} \\
\bottomrule
\end{tabular}%
}
\end{table}

\paragraph{No-retrieval control:} Consistent with $G^2$-Reader, ten text chunks and ten visual elements are deterministically sampled from the candidate documents and interleaved with the question in a chain-of-thought prompt (Appendix~\ref{app:singlevlm}). 

\paragraph{Published systems:} The primary baseline is $G^2$-Reader~\citep{du2026g}, which represents the previous state of the art on VisDoMBench. Additional comparisons are made with recent multimodal RAG and document question answering systems: GPT-5~\citep{singh2025openai}, Deepseek-OCR~\citep{wei2025deepseek}, RAGAnything~\citep{guo2025rag}, MA-RAG~\citep{nguyen2025ma}, GraphRAG~\citep{edge2024local}, LightRAG~\citep{guo2024lightrag}, MMGraphRAG~\citep{wan2025mmgraphrag}, VisDoMRAG~\citep{suri2025visdom}, and ViDoRAG~\citep{wang2025vidorag}. Their scores are cited from the $G^2$-Reader paper, which evaluates all systems under the same candidate-set protocol, evidence budget, and LLM-as-judge pipeline adopted in this study.
\begin{table*}[t]
\centering
\scriptsize
\setlength{\tabcolsep}{5pt}
\caption{Comparison with alternative selection strategies. Accuracy (\%),
mean\,$\pm$\,std over 3 runs; Qwen3-VL-32B reader. MMR uses \(\lambda{=}0.5\),
DPP uses \(\theta{=}1\). \textbf{Avg.}\ is the VisDoMBench average; MMLongBench-Doc
is a separate benchmark. CDS is best on every dataset.}
\label{tab:selectors}
\resizebox{\textwidth}{!}{%
\begin{tabular}{l ccccc >{\columncolor{gray!15}}c !{\vrule width 0.8pt} c}
\toprule
Method & FetaTab & PaperTab & SciGraphQA & SlideVQA & SPIQA & \textbf{Avg.} & MMLongBench \\
\midrule
MMR ($\lambda{=}0.5$) & 57.0$\pm$0.0 & 55.5$\pm$0.7 & 44.7$\pm$0.2 & 46.3$\pm$0.4 & 69.9$\pm$0.5 & 54.7$\pm$0.2 & 42.8$\pm$0.2 \\
DPP & 44.1$\pm$0.2 & 45.6$\pm$0.5 & 34.4$\pm$0.3 & 35.6$\pm$0.6 & 62.5$\pm$0.4 & 44.4$\pm$0.3 & 41.2$\pm$0.1 \\
PPR & 59.8$\pm$0.2 & 52.4$\pm$0.9 & 49.1$\pm$0.7 & 54.8$\pm$0.2 & 55.9$\pm$0.4 & 54.4$\pm$0.1 & 40.4$\pm$0.3 \\
\textbf{CDS} & \textbf{71.0$\pm$0.6} & \textbf{65.6$\pm$0.4} & \textbf{57.4$\pm$0.4} & \textbf{62.2$\pm$0.4} & \textbf{78.8$\pm$0.2} & \textbf{67.0$\pm$0.2} & \textbf{45.0$\pm$0.4} \\
\bottomrule
\end{tabular}}
\end{table*}

\paragraph{Selection-rule baselines.} CDS is evaluated against three selection rules using the same per-query graph and encoder, ensuring that only the selection mechanism differs. Full algorithmic details, including the exact MMR, $k$-DPP, and PPR updates, are deferred to Appendix~\ref{app:baselines}.   \textbf{MMR}~\citep{carbonell1998use} greedily maximizes $\lambda\cos(z_q, z_i) - (1-\lambda)\max_{j\in S}\cos(z_i, z_j)$; $\lambda$ is swept over $\{0.3, 0.5, 0.7, 0.9\}$, and the optimal per-dataset setting is reported. \textbf{$k$-DPP}~\citep{kulesza2011k} selects a size-$B$ subset by greedy MAP on $L_{ij} = q_i q_j K_{ij}$, where $q_i = \max(\cos(z_q, z_i), 0)$ and $K_{ij} = \cos(z_i, z_j)$, thereby promoting diversity through determinantal volume rather than query-anchored repulsion. Personalized PageRank (PPR) as a retriever returns the top-$B$ notes by Personalized PageRank from a query-seeded teleport, evaluating whether graph structure alone is sufficient. These baselines isolate the relevance/diversity trade-off, the diversification mechanism, and the graph structure. Any remaining performance gap can thus be attributed to CDS's joint use of similarity, repulsion, and the query-as-constraint formulation. For direct comparison with $G^2$-Reader, plain cosine top-$B$ retrieval is reported on a 250-sample subset (Table~\ref{tab:sample250}) as the relevance-only reference on identical data.

\section{Experimental Setup}
We evaluate on \textbf{VisDoMBench}~\citep{suri2025visdom} and \textbf{MMLongBench-Doc}~\citep{ma2024mmlongbench}, following the $G^2$-Reader protocol~\citep{du2026g}: each query is answered against a five-document candidate pool (gold $+$ four distractors), and answers are scored by a GPT-4o-mini judge~\citep{zheng2023judging}. Unless noted, CDS uses an evidence budget $B{=}10$, no query-row sparsification, and uniform initialization; the reader is \textbf{Qwen3-VL-32B}~\citep{bai2025qwen3}, with \textbf{Qwen2.5-VL-7B}~\citep{bai2025qwen2} and \textbf{GLM-4.1V-9B}~\cite{hong2025glm} reported for scaling. 
\begin{figure*}
    \centering
    \includegraphics[width=0.7\linewidth]{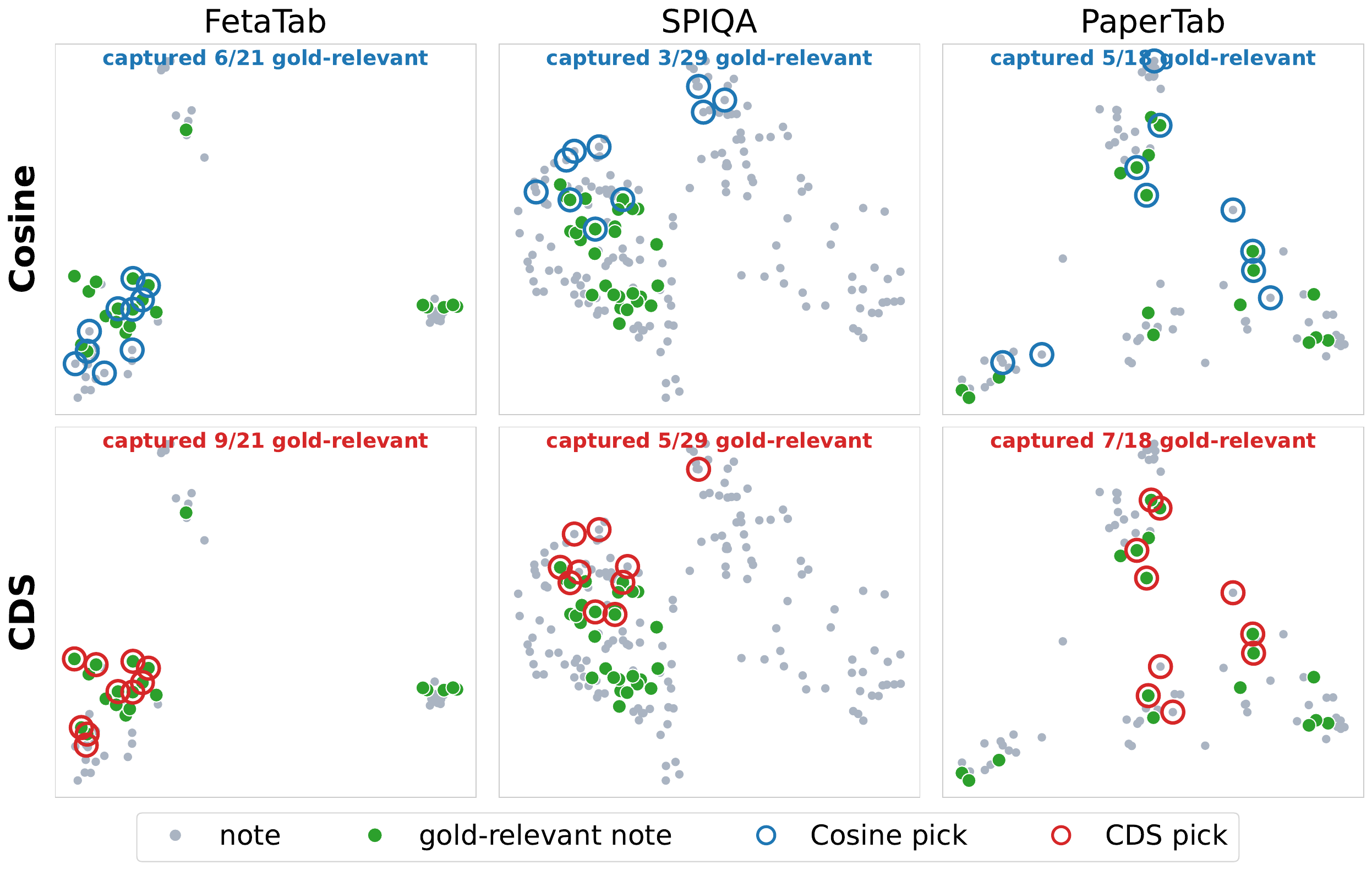}
    \caption{Cosine vs.\ CDS evidence selection on the per-query note map (2-D Principal Component Analysis, PCA); grey = note, green = gold-relevant). Top: cosine top-$B$; bottom: CDS. Cosine concentrates its picks in a single region and misses much of the gold evidence, whereas CDS spreads its selection and captures more gold-relevant notes (counts shown per panel). Panels are queries where CDS most outperforms cosine in coverage.}
    \label{fig:cosine_vs_cds}
\end{figure*}
\subsection{Main Results}
\label{sec:main_results}
Table~\ref{tab:main_results} presents results for the full VisDoMBench. The proposed method, \textbf{Qwen3-VL-32B+CDS}, achieves the highest average accuracy (\textbf{66.99}), surpassing all previous systems, including the strongest graph- and agent-based approaches: $G^2$-Reader (66.21) and VisDoMRAG (65.01). This performance is attained using a single replicator-dynamics retrieval step rather than multi-agent planning or explicit document-graph construction. CDS ranks first on SPIQA (78.85) and PaperTab (65.59), and second on FetaTab (70.90). However, it is less competitive on SciGraphQA and SlideVQA, where chart- and slide-specialized methods such as LightRAG and ViDoRAG perform better. The impact of retrieval is considerable: compared to the no-retrieval Qwen3-VL-32B baseline (29.90), CDS provides an average improvement of $+37.1$ points. This result suggests that evidence selection, rather than reader capacity, constitutes the primary bottleneck in these multi-document tasks.

Table~\ref{tab:mmlb} presents MMLongBench-Doc results for three readers of increasing scale. CDS outperforms the no-retrieval baseline for Qwen3-VL-32B ($40.19\to45.01$, $+4.82$) and Qwen2.5-VL-7B ($28.36\to32.30$, $+3.94$), while performing slightly below the baseline for GLM-4.1V ($41.04\to39.15$). The observed gains are smaller than those on VisDoMBench, as MMLongBench-Doc questions are based on a single long document, where directly providing sampled content already constitutes a strong baseline. Nevertheless, CDS benefits both Qwen readers and remains competitive with GLM, demonstrating that the approach generalizes to the long-single-document setting without modification. Importantly, these results are achieved using a lightweight open text encoder (\texttt{nomic-embed-text-v1.5}, 768-d), whereas competing systems employ larger embeddings (e.g., OpenAI \texttt{text-embedding-3-small}). Achieving comparable or superior performance with a less powerful retrieval backbone highlights that the improvements are attributable to the selection formulation rather than the embedding model.

On the identical subset used in $G^2$-Reader's ablation, evaluated with the same reader and judge and differing only in the selection rule, CDS outperforms a plain cosine top-$B$ retriever by an average of $+4.0$ points ($69.5$ vs. $65.5$, Table~\ref{tab:sample250}). CDS achieves gains on every dataset except PaperTab, where performance is tied, and surpasses $G^2$-Reader's reported $68.0$ on the same data~\citep{du2026g}, despite replacing agentic planning with a single, training-free retrieval step. Figure~\ref{fig:cosine_vs_cds} illustrates this distinction. The cosine method concentrates selections within a single embedding region, whereas CDS distributes selections across the gold-relevant subspace, thereby capturing significantly more gold notes per query.

\paragraph{Comparison with alternative selection strategies.} The necessity of the CDS formulation is evaluated against simpler relevance-plus-diversity selectors. Table~\ref{tab:selectors} presents a comparison of CDS with MMR ($\lambda=0.5$), $k$-DPP, and PPR-as-retriever, as defined in Section~\ref{sec:baselines}. CDS achieves the highest performance across all datasets and, on average ($67.0$), outperforms MMR ($54.7$), DPP ($44.4$), and PPR ($54.4$). The two diversity-based selectors underperform because, at a balanced relevance and diversity weighting, they tend to include off-topic but dissimilar notes in the evidence set, and performance improves only when diversity is entirely suppressed. PPR does not perform better; relying solely on graph structure provides a weaker signal than embedding relevance. CDS avoids this limitation because the spectral bound on its constraint parameter automatically balances query relevance with inter-note dissimilarity, enabling productive diversification without manual trade-off tuning or per-dataset adjustment.

\paragraph{Additional Experiments:} Appendix~\ref{app:additional} presents additional ablation studies on Qwen3-VL-32B. The initialization prior (Table~\ref{tab:prior}) and graph-evolution iterations (Table~\ref{tab:evolution}) demonstrate that CDS is robust to both seeding and refinement choices. Query-row sparsification (Appendix~\ref{app:tau}, Table~\ref{tab:tau}) indicates that performance remains consistent for $\tau \in [0, 0.3]$ and declines only with aggressive thresholding. The diversification-block sign ablation (Appendix~\ref{app:divblock}, Table~\ref{tab:divblock}) reveals that converting note--note edges from dissimilarity to similarity reduces the VisDoMBench average by 4.0 points, confirming that dissimilarity coupling underpins CDS's advantage. Finally, the evidence-budget sweep (Appendix~\ref{app:evibud}, Table~\ref{tab:budget}) and diversification-strength sweep (Table~\ref{tab:beta}) demonstrate that accuracy remains stable across $B \in [10, 20]$ and $\beta \in [0.5, 1]$

\section{Conclusion}
This work formulates multimodal evidence selection for document question answering as a CDS problem on a query-augmented graph. A single query (constraint) vertex contributes relevance via its cosine edges, while a node-to-node dissimilarity block encourages non-redundancy. Replicator dynamics yield a soft membership whose support constitutes a relevant and diverse evidence set. The method is training-free, requires no tuning of a relevance-diversity trade-off parameter, and employs a parameter-free uniform initialization. On the VisDoMBench benchmark, the approach achieves the highest average accuracy ($66.99$), outperforming advanced graph- and agent-based systems such as $G^2$-Reader and VisDoMRAG, despite its simplicity. It also surpasses a no-retrieval baseline by a substantial margin, indicating that evidence selection, rather than reader capacity, is the primary bottleneck in these multi-document tasks. Extensive ablations (Appendix~\ref{app:additional}) demonstrate the robustness of the approach. These findings suggest that constraint-driven graph clustering warrants further investigation as a general retrieval primitive. Future directions include open-corpus retrieval with larger candidate pools, multi-query and pinned-memory scenarios already supported by the formulation, and closer integration between memory-graph evolution and the selection objective.
\section*{Limitations}
The proposed method is evaluated using the same five-document retrieval protocol (gold plus four distractors) as $G^2$-Reader and previous VisDoMBench studies, which ensures a fair and controlled comparison across methods. Nevertheless, this protocol constitutes a relatively constrained retrieval setting. The advantages of diversity-aware selection are expected to become more evident under more stringent conditions, such as when candidate pools are larger or when distractors are more challenging. Stress tests involving larger candidate pools, more difficult distractors, or open-corpus retrieval are reserved for future research. Performance improvements are also less substantial in the single-long-document setting (MMLongBench-Doc), where selection is less critical than in the multi-document scenario for which CDS is primarily designed. The current evaluation includes two benchmarks, three open VLM readers, and an LLM-based judge. Expanding to broader domains, incorporating additional readers, and conducting human evaluations are identified as promising directions for future investigation. Furthermore, CDS does not uniformly benefit all reader VLMs: on MMLongBench-Doc, GLM-4.1V-9B performs marginally below its no-retrieval baseline (-1.89 points), suggesting that the interaction between retrieved-evidence formatting and reader-specific instruction-following warrants further investigation.

\section*{Acknowledgements}
This work was supported by the European Union's Horizon Europe research and
innovation programme under the Marie Sk\l{}odowska-Curie grant agreement
No.~101205348 (CASPER). We acknowledge the EuroHPC Joint Undertaking for
awarding this project access to the EuroHPC supercomputer LEONARDO, hosted by
CINECA (Italy) and the LEONARDO consortium, through the EuroHPC AI Factories
``AI for Science and Collaborative EU Projects'' Access call (proposal
No.~EHPC-AIF-2026SC01-041). We further acknowledge the CINECA award under the
ISCRA initiative (Class C project IsCd5\_CASPER-A), for the availability of
high performance computing resources and support. Views and opinions expressed
are however those of the author(s) only and do not necessarily reflect those of
the European Union or the European Research Executive Agency. Neither the
European Union nor the granting authority can be held responsible for them.



\bibliographystyle{unsrtnat}
\bibliography{custom}

@article{chen2018fast,
  title={Fast greedy map inference for determinantal point process to improve recommendation diversity},
  author={Chen, Laming and Zhang, Guoxin and Zhou, Eric},
  journal={Advances in neural information processing systems},
  volume={31},
  year={2018}
}

@article{kulesza2012determinantal,
  title={Determinantal point processes for machine learning},
  author={Kulesza, Alex and Taskar, Ben},
  journal={Foundations and Trends{\textregistered} in Machine Learning},
  volume={5},
  number={2-3},
  pages={123--286},
  year={2012},
  publisher={Emerald Publishing Limited}
}

@article{du2026g,
  title={{{$g^2$-Reader: Dual Evolving Graphs for Multimodal Document Comprehension}}},
  author={Du, Yaxin and Song, Junru and Zhou, Yifan and Wang, Cheng and Gu, Jiahao and Chen, Zimeng and Chen, Menglan and Yao, Wen and Yang, Yang and Wen, Ying and others},
  journal={arXiv preprint arXiv:2601.22055},
  year={2026}
}

@inproceedings{kulesza2011k,
  title={k-dpps: Fixed-size determinantal point processes},
  author={Kulesza, Alex and Taskar, Ben},
  booktitle={Proceedings of the 28th International Conference on Machine Learning (ICML-11)},
  pages={1193--1200},
  year={2011}
}

@article{gillenwater2012near,
  title={Near-optimal map inference for determinantal point processes},
  author={Gillenwater, Jennifer and Kulesza, Alex and Taskar, Ben},
  journal={Advances in Neural Information Processing Systems},
  volume={25},
  year={2012}
}

@article{pavan2007dominant,
  title={Dominant sets and pairwise clustering},
  author={Pavan, Massimiliano and Pelillo, Marcello},
  journal={IEEE transactions on pattern analysis and machine intelligence},
  volume={29},
  number={1},
  pages={167--172},
  year={2007},
  publisher={IEEE}
}

@article{bulo2017dominant,
  title={Dominant-set clustering: A review},
  author={Bul{\`o}, Samuel Rota and Pelillo, Marcello},
  journal={European Journal of Operational Research},
  volume={262},
  number={1},
  pages={1--13},
  year={2017},
  publisher={Elsevier}
}

@article{zemene2018dominant,
  title={Dominant sets for “constrained” image segmentation},
  author={Zemene, Eyasu Zemene and Alemu, Leulseged Tesfaye and Pelillo, Marcello},
  journal={IEEE Transactions on Pattern Analysis and Machine Intelligence},
  volume={41},
  number={10},
  pages={2438--2451},
  year={2018},
  publisher={IEEE}
}

@article{tesfaye2019multi,
  title={Multi-target tracking in multiple non-overlapping cameras using fast-constrained dominant sets},
  author={Tesfaye, Yonatan Tariku and Zemene, Eyasu and Prati, Andrea and Pelillo, Marcello and Shah, Mubarak},
  journal={International Journal of Computer Vision},
  volume={127},
  number={9},
  pages={1303--1320},
  year={2019},
  publisher={Springer}
}

@article{pelillo1998replicator,
  title={Replicator equations, maximal cliques, and graph isomorphism},
  author={Pelillo, Marcello},
  journal={Advances in Neural Information Processing Systems},
  volume={11},
  year={1998}
}

@article{bomze1997evolution,
  title={Evolution towards the maximum clique},
  author={Bomze, Immanuel M},
  journal={Journal of Global Optimization},
  volume={10},
  number={2},
  pages={143--164},
  year={1997},
  publisher={Springer}
}

@article{packer2023memgpt,
  title={MemGPT: Towards LLMs as operating systems},
  author={Packer, Charles and Fang, Vivian and Patil, Shishir G and Lin, Kevin and Wooders, Sarah and Gonzalez, Joseph E},
  journal={arXiv preprint arXiv:2310.08560},
  year={2023}
}

@inproceedings{zhong2024memorybank,
  title={Memorybank: Enhancing large language models with long-term memory},
  author={Zhong, Wanjun and Guo, Lianghong and Gao, Qiqi and Ye, He and Wang, Yanlin},
  booktitle={Proceedings of the AAAI conference on artificial intelligence},
  volume={38},
  pages={19724--19731},
  year={2024}
}

@article{xu2026mem,
  title={A-mem: Agentic memory for llm agents},
  author={Xu, Wujiang and Liang, Zujie and Mei, Kai and Gao, Hang and Tan, Juntao and Zhang, Yongfeng},
  journal={Advances in Neural Information Processing Systems},
  volume={38},
  pages={17577--17604},
  year={2026}
}

@article{hong2025glm,
  title={Glm-4.1 v-thinking: Towards versatile multimodal reasoning with scalable reinforcement learning},
  author={Hong, Wenyi and Yu, Wenmeng and Gu, Xiaotao and Wang, Guo and Gan, Guobing and Tang, Haomiao and Cheng, Jiale and Qi, Ji and Ji, Junhui and Pan, Lihang and others},
  journal={arXiv e-prints},
  pages={arXiv--2507},
  year={2025}
}

@article{wei2025deepseek,
  title={Deepseek-ocr: Contexts optical compression},
  author={Wei, Haoran and Sun, Yaofeng and Li, Yukun},
  journal={arXiv preprint arXiv:2510.18234},
  year={2025}
}

@article{singh2025openai,
  title={Openai gpt-5 system card},
  author={Singh, Aaditya and Fry, Adam and Perelman, Adam and Tart, Adam and Ganesh, Adi and El-Kishky, Ahmed and McLaughlin, Aidan and Low, Aiden and Ostrow, AJ and Ananthram, Akhila and others},
  journal={arXiv preprint arXiv:2601.03267},
  year={2025}
}

@article{guo2025rag,
  title={Rag-anything: All-in-one rag framework},
  author={Guo, Zirui and Ren, Xubin and Xu, Lingrui and Zhang, Jiahao and Huang, Chao},
  journal={arXiv preprint arXiv:2510.12323},
  year={2025}
}

@article{nguyen2025ma,
  title={Ma-rag: Multi-agent retrieval-augmented generation via collaborative chain-of-thought reasoning},
  author={Nguyen, Thang and Chin, Peter and Tai, Yu-Wing},
  journal={arXiv preprint arXiv:2505.20096},
  year={2025}
}

@article{edge2024local,
  title={From local to global: A graph rag approach to query-focused summarization},
  author={Edge, Darren and Trinh, Ha and Cheng, Newman and Bradley, Joshua and Chao, Alex and Mody, Apurva and Truitt, Steven and Metropolitansky, Dasha and Ness, Robert Osazuwa and Larson, Jonathan},
  journal={arXiv preprint arXiv:2404.16130},
  year={2024}
}

@article{guo2024lightrag,
  title={Lightrag: Simple and fast retrieval-augmented generation},
  author={Guo, Zirui and Xia, Lianghao and Yu, Yanhua and Ao, Tian and Huang, Chao},
  journal={arXiv preprint arXiv:2410.05779},
  volume={2},
  number={3},
  year={2024}
}

@article{wan2025mmgraphrag,
  title={Mmgraphrag: Bridging vision and language with interpretable multimodal knowledge graphs},
  author={Wan, Xueyao and Yu, Hang},
  journal={arXiv preprint arXiv:2507.20804},
  year={2025}
}

@inproceedings{suri2025visdom,
  title={Visdom: Multi-document qa with visually rich elements using multimodal retrieval-augmented generation},
  author={Suri, Manan and Mathur, Puneet and Dernoncourt, Franck and Goswami, Kanika and Rossi, Ryan A and Manocha, Dinesh},
  booktitle={Proceedings of the 2025 Conference of the Nations of the Americas Chapter of the Association for Computational Linguistics: Human Language Technologies (Volume 1: Long Papers)},
  pages={6088--6109},
  year={2025}
}

@inproceedings{wang2025vidorag,
  title={Vidorag: Visual document retrieval-augmented generation via dynamic iterative reasoning agents},
  author={Wang, Qiuchen and Ding, Ruixue and Chen, Zehui and Wu, Weiqi and Wang, Shihang and Xie, Pengjun and Zhao, Feng},
  booktitle={Proceedings of the 2025 Conference on Empirical Methods in Natural Language Processing},
  pages={9124--9145},
  year={2025}
}

@article{ma2024mmlongbench,
  title={Mmlongbench-doc: Benchmarking long-context document understanding with visualizations},
  author={Ma, Yubo and Zang, Yuhang and Chen, Liangyu and Chen, Meiqi and Jiao, Yizhu and Li, Xinze and Lu, Xinyuan and Liu, Ziyu and Ma, Yan and Dong, Xiaoyi and others},
  journal={Advances in Neural Information Processing Systems},
  volume={37},
  pages={95963--96010},
  year={2024}
}

@article{liu2024lost,
  title={Lost in the middle: How language models use long contexts},
  author={Liu, Nelson F and Lin, Kevin and Hewitt, John and Paranjape, Ashwin and Bevilacqua, Michele and Petroni, Fabio and Liang, Percy},
  journal={Transactions of the association for computational linguistics},
  volume={12},
  pages={157--173},
  year={2024}
}

@article{faysse2024colpali,
  title={Colpali: Efficient document retrieval with vision language models},
  author={Faysse, Manuel and Sibille, Hugues and Wu, Tony and Omrani, Bilel and Viaud, Gautier and Hudelot, C{\'e}line and Colombo, Pierre},
  journal={arXiv preprint arXiv:2407.01449},
  year={2024}
}

@inproceedings{mathew2021docvqa,
  title={Docvqa: A dataset for vqa on document images},
  author={Mathew, Minesh and Karatzas, Dimosthenis and Jawahar, CV},
  booktitle={Proceedings of the IEEE/CVF winter conference on applications of computer vision},
  pages={2200--2209},
  year={2021}
}

@article{cho2024m3docrag,
  title={M3docrag: Multi-modal retrieval is what you need for multi-page multi-document understanding},
  author={Cho, Jaemin and Mahata, Debanjan and Irsoy, Ozan and He, Yujie and Bansal, Mohit},
  journal={arXiv preprint arXiv:2411.04952},
  year={2024}
}

@inproceedings{tanaka2023slidevqa,
  title={Slidevqa: A dataset for document visual question answering on multiple images},
  author={Tanaka, Ryota and Nishida, Kyosuke and Nishida, Kosuke and Hasegawa, Taku and Saito, Itsumi and Saito, Kuniko},
  booktitle={Proceedings of the AAAI Conference on Artificial Intelligence},
  volume={37},
  pages={13636--13645},
  year={2023}
}

@article{pramanick2024spiqa,
  title={Spiqa: A dataset for multimodal question answering on scientific papers},
  author={Pramanick, Shraman and Chellappa, Rama and Venugopalan, Subhashini},
  journal={Advances in Neural Information Processing Systems},
  volume={37},
  pages={118807--118833},
  year={2024}
}

@article{li2023scigraphqa,
  title={Scigraphqa: A large-scale synthetic multi-turn question-answering dataset for scientific graphs},
  author={Li, Shengzhi and Tajbakhsh, Nima},
  journal={arXiv preprint arXiv:2308.03349},
  year={2023}
}

@article{hui2024uda,
  title={Uda: A benchmark suite for retrieval augmented generation in real-world document analysis},
  author={Hui, Yulong and Lu, Yao and Zhang, Huanchen},
  journal={Advances in Neural Information Processing Systems},
  volume={37},
  pages={67200--67217},
  year={2024}
}

@inproceedings{deng2025longdocurl,
  title={Longdocurl: a comprehensive multimodal long document benchmark integrating understanding, reasoning, and locating},
  author={Deng, Chao and Yuan, Jiale and Bu, Pi and Wang, Peijie and Li, Zhong-Zhi and Xu, Jian and Li, Xiao-Hui and Gao, Yuan and Song, Jun and Zheng, Bo and others},
  booktitle={Proceedings of the 63rd Annual Meeting of the Association for Computational Linguistics (Volume 1: Long Papers)},
  pages={1135--1159},
  year={2025}
}

@article{tito2023hierarchical,
  title={Hierarchical multimodal transformers for multipage docvqa},
  author={Tito, Rub{\`e}n and Karatzas, Dimosthenis and Valveny, Ernest},
  journal={Pattern Recognition},
  volume={144},
  pages={109834},
  year={2023},
  publisher={Elsevier}
}

@inproceedings{carbonell1998use,
  title={The use of MMR, diversity-based reranking for reordering documents and producing summaries},
  author={Carbonell, Jaime and Goldstein, Jade},
  booktitle={Proceedings of the 21st annual international ACM SIGIR conference on Research and development in information retrieval},
  pages={335--336},
  year={1998}
}

@article{zemene2018large,
  title={Large-scale image geo-localization using dominant sets},
  author={Zemene, Eyasu and Tesfaye, Yonatan Tariku and Idrees, Haroon and Prati, Andrea and Pelillo, Marcello and Shah, Mubarak},
  journal={IEEE transactions on pattern analysis and machine intelligence},
  volume={41},
  number={1},
  pages={148--161},
  year={2018},
  publisher={IEEE}
}

@article{ALEMU2020103862,
title = {Multi-feature fusion for image retrieval using constrained dominant sets},
journal = {Image and Vision Computing},
volume = {94},
pages = {103862},
year = {2020},
issn = {0262-8856},
doi = {https://doi.org/10.1016/j.imavis.2019.103862},
url = {https://www.sciencedirect.com/science/article/pii/S026288561930455X},
author = {Leulseged Tesfaye Alemu and Marcello Pelillo}
}

@inproceedings{alemu2019deep,
  title={Deep constrained dominant sets for person re-identification},
  author={Alemu, Leulseged Tesfaye and Pelillo, Marcello and Shah, Mubarak},
  booktitle={Proceedings of the IEEE/CVF international conference on computer vision},
  pages={9855--9864},
  year={2019}
}

@article{wang2024mineru,
  title={Mineru: An open-source solution for precise document content extraction},
  author={Wang, Bin and Xu, Chao and Zhao, Xiaomeng and Ouyang, Linke and Wu, Fan and Zhao, Zhiyuan and Xu, Rui and Liu, Kaiwen and Qu, Yuan and Shang, Fukai and others},
  journal={arXiv preprint arXiv:2409.18839},
  year={2024}
}

@inproceedings{zemene2016interactive,
  title={Interactive image segmentation using constrained dominant sets},
  author={Zemene, Eyasu and Pelillo, Marcello},
  booktitle={European Conference on Computer Vision},
  pages={278--294},
  year={2016},
  organization={Springer}
}

@article{lewis2020retrieval,
  title={Retrieval-augmented generation for knowledge-intensive nlp tasks},
  author={Lewis, Patrick and Perez, Ethan and Piktus, Aleksandra and Petroni, Fabio and Karpukhin, Vladimir and Goyal, Naman and K{\"u}ttler, Heinrich and Lewis, Mike and Yih, Wen-tau and Rockt{\"a}schel, Tim and others},
  journal={Advances in neural information processing systems},
  volume={33},
  pages={9459--9474},
  year={2020}
}

@article{nan2022fetaqa,
  title={FeTaQA: Free-form table question answering},
  author={Nan, Linyong and Hsieh, Chiachun and Mao, Ziming and Lin, Xi Victoria and Verma, Neha and Zhang, Rui and Kry{\'s}ci{\'n}ski, Wojciech and Schoelkopf, Hailey and Kong, Riley and Tang, Xiangru and others},
  journal={Transactions of the Association for Computational Linguistics},
  volume={10},
  pages={35--49},
  year={2022},
  publisher={MIT Press One Broadway, 12th Floor, Cambridge, Massachusetts 02142, USA~…}
}

@inproceedings{chia2025m,
  title={M-longdoc: A benchmark for multimodal super-long document understanding and a retrieval-aware tuning framework},
  author={Chia, Yew Ken and Cheng, Liying and Chan, Hou Pong and Song, Maojia and Liu, Chaoqun and Aljunied, Mahani and Poria, Soujanya and Bing, Lidong},
  booktitle={Proceedings of the 2025 Conference on Empirical Methods in Natural Language Processing},
  pages={9244--9261},
  year={2025}
}

@article{bai2025qwen3,
  title={Qwen3-vl technical report},
  author={Bai, Shuai and Cai, Yuxuan and Chen, Ruizhe and Chen, Keqin and Chen, Xionghui and Cheng, Zesen and Deng, Lianghao and Ding, Wei and Gao, Chang and Ge, Chunjiang and others},
  journal={arXiv preprint arXiv:2511.21631},
  year={2025}
}

@article{bai2025qwen2,
  title={Qwen2. 5-VL Technical Report},
  author={Bai, Shuai and Chen, Keqin and Liu, Xuejing and Wang, Jialin and Ge, Wenbin and Song, Sibo and Dang, Kai and Wang, Peng and Wang, Shijie and Tang, Jun and others},
  journal={arXiv preprint arXiv:2502.13923},
  year={2025}
}

@article{nussbaum2024nomic,
  title={Nomic embed: Training a reproducible long context text embedder},
  author={Nussbaum, Zach and Morris, John X and Duderstadt, Brandon and Mulyar, Andriy},
  journal={arXiv preprint arXiv:2402.01613},
  year={2024}
}

@inproceedings{kwon2023efficient,
  title={Efficient memory management for large language model serving with pagedattention},
  author={Kwon, Woosuk and Li, Zhuohan and Zhuang, Siyuan and Sheng, Ying and Zheng, Lianmin and Yu, Cody Hao and Gonzalez, Joseph and Zhang, Hao and Stoica, Ion},
  booktitle={Proceedings of the 29th symposium on operating systems principles},
  pages={611--626},
  year={2023}
}

@inproceedings{ma2024unifying,
  title={Unifying multimodal retrieval via document screenshot embedding},
  author={Ma, Xueguang and Lin, Sheng-Chieh and Li, Minghan and Chen, Wenhu and Lin, Jimmy},
  booktitle={Proceedings of the 2024 Conference on Empirical Methods in Natural Language Processing},
  pages={6492--6505},
  year={2024}
}

@article{zheng2023judging,
  title={Judging llm-as-a-judge with mt-bench and chatbot arena},
  author={Zheng, Lianmin and Chiang, Wei-Lin and Sheng, Ying and Zhuang, Siyuan and Wu, Zhanghao and Zhuang, Yonghao and Lin, Zi and Li, Zhuohan and Li, Dacheng and Xing, Eric and others},
  journal={Advances in neural information processing systems},
  volume={36},
  pages={46595--46623},
  year={2023}
}

\appendix
\newpage
\section*{The Use of Large Language Models}
We employed a Large Language Model to assist with reducing wordy paragraphs to help the paper fit within the page limit.
\section{Algorithm: Pseudocode}
\label{app:algorithm}
Algorithm~\ref{alg:cds-retrieval} summarises the end-to-end CDS retrieval procedure described in Section~\ref{sec:cds-graph}, from raw query and document nodes to the final reader output. The algorithm has three logical stages. Lines 1--2 construct the query-augmented affinity matrix $A$: note--note entries encode dissimilarity ($1-\mathbf{z}_i^\top\mathbf{z}_j$) so that redundant evidence repels, while query--note entries encode similarity ($\max(\mathbf{z}_q^\top\mathbf{z}_i, 0)$) so that the query attracts evidence aligned with the question. Line 3--5 optionally apply the sparsification threshold $\tau$ to prune weak query edges (Appendix~\ref{app:tau}). Line 6 forms the CDS payoff matrix $M$ by subtracting $\alpha\mathbf{I}_{\bar{\mathcal{S}}}$ from $A$ on the complement of the constraint set $\mathcal{S}=\{q\}$, where $\alpha = \lambda_{\max}(A) + \varepsilon$ is the spectral bound that guarantees the query vertex appears in the dominant cluster. Line 7 runs the replicator dynamics from the initial prior $\pi$ (uniform by default; see Appendix~\ref{app:additional}) until convergence to the local maximiser $\mathbf{x}^\star$ of $\mathbf{x}^\top M\mathbf{x}$ on the simplex. The shift constant $c$ ensures the multiplicative update is well-defined when $M$ has negative entries. Line 8 reads off the evidence set $R$ as the top-$B$ document vertices by $x_i^\star$ above the support threshold $\theta$, discarding the query vertex. Line 9 hands $(q, R)$ to the VLM reader for grounded answer generation. The procedure is deterministic given the inputs and requires no training, no relevance--diversity trade-off parameter, and no greedy locking; selection is the equilibrium of a single quadratic program on the query-augmented graph.
\begin{algorithm}[t]
\caption{CDS: Multimodal evidence retrieval}
\label{alg:cds-retrieval}
\footnotesize
\begin{algorithmic}[1]
\Statex \textbf{Input:} document nodes $\mathcal{V}=\{1,\dots,N\}$ with $\ell_2$-normalised
embeddings $\{\mathbf{z}_i\}$; question $q$; evidence budget $B$; sparsity threshold
$\tau$; init prior $\boldsymbol{\pi}$ (uniform by default); embedder $f_{\mathrm{emb}}$; VLM $f_{\mathrm{VLM}}$.
\Statex \textbf{Output:} answer $\hat{a}$.

\State $\mathbf{z}_q \gets \mathrm{norm}\big(f_{\mathrm{emb}}(q)\big)$
\Statex \textit{Vertices: $N$ notes and one query (anchor) vertex; constraint set
$S=\{q\}$, complement $\bar{S}=\mathcal{V}$.}

\State \textbf{Build affinity} $A\in\mathbb{R}^{(N+1)\times(N+1)}$ from embeddings:
\Statex \quad $A_{ij}=1-\mathbf{z}_i^{\top}\mathbf{z}_j,\;\;
A_{qi}=A_{iq}=\max(\mathbf{z}_q^{\top}\mathbf{z}_i,\,0),\;\;
\operatorname{diag}(A)=0$
\Statex \quad\Comment{notes repel one another (dissimilarity); query attracts similar notes}

\If{$\tau>0$}
  \State zero query edges with $\mathbf{z}_q^{\top}\mathbf{z}_i<\tau$ (keep the single strongest if none clear $\tau$)
\EndIf

\State \textbf{CDS payoff:}\;
$\alpha \gets \lambda_{\max}(A)+\epsilon,\quad M \gets A-\alpha\,I_{\bar{S}}$
\Statex \quad\Comment{$M_{qq}=0$; $M_{ii}=-\alpha$ for notes $\Rightarrow$ support must contain $S$}

\State \textbf{Solve} by replicator dynamics ($\mathbf{x}^{(0)}=\boldsymbol{\pi}$,
shift $c=\max(0,-\min M)+\varepsilon$):
\Statex \quad $\displaystyle
\mathbf{x}^{(t+1)}=
\frac{\mathbf{x}^{(t)}\odot\big(M\mathbf{x}^{(t)}+c\big)}
     {\big(\mathbf{x}^{(t)}\big)^{\top}M\mathbf{x}^{(t)}+c},
\qquad
\mathbf{x}^{\ast}=\lim_{t\to\infty}\mathbf{x}^{(t)}\in\Delta$
\Statex \quad\Comment{local maximiser of $\mathbf{x}^{\top}M\mathbf{x}$ on the simplex (constrained dominant set)}

\State \textbf{Select evidence:}\;
$R \gets \operatorname{Top\text{-}}B\{\,x^{\ast}_i : i\in\mathcal{V},\; x^{\ast}_i>\theta\,\}$
\Statex \quad\Comment{notes only, ranked by $x^{\ast}_i$; anchor excluded}

\State \textbf{Generate:}\; $\hat{a}\gets f_{\mathrm{VLM}}(q,\,R)$
\State \Return $\hat{a}$
\end{algorithmic}
\end{algorithm}

\section{Baseline Implementation Details}
\label{app:baselines}

All retrieval baselines operate on the same per-query memory graph and the same text encoder as CDS, so that any difference is attributable to the selection rule alone. Let the graph contain $N$ notes with $\ell_2$-normalized embeddings $\mathbf{z}_1,\dots,\mathbf{z}_N\in\mathbb{R}^{d}$ ($d{=}768$, \texttt{nomic-embed-text-v1.5}) and let $\mathbf{z}_q$ be the normalized query embedding. We define the query relevance of note $i$ and the note--note similarity as:

\begin{equation}
\mathrm{rel}_i = \mathbf{z}_q^{\top}\mathbf{z}_i, \qquad S_{ij} = \mathbf{z}_i^{\top}\mathbf{z}_j ,
\label{eq:rel-sim}
\end{equation}

where $S=EE^{\top}$ (with $E$ the stacked embeddings) is symmetric positive-semidefinite and $S_{ii}=1$. Every method returns an evidence set $R$ of size $B$ (the budget; $B{=}10$ unless stated), which is passed to the reader.

\subsection{Cosine Retrieval}
The simplest baseline ranks notes by query relevance and skips all graph dynamics:

\begin{equation}
R = \operatorname*{Top\text{-}}{B}_{\,i\in\{1,\dots,N\}}\ \mathrm{rel}_i .
\label{eq:cosine}
\end{equation}

This isolates whether the replicator dynamics of CDS add value over plain similarity ranking.

\subsection{Random Selection (control)}
A no-retrieval control that keeps the reader and budget fixed but removes any notion of relevance: $R$ is drawn uniformly without replacement from the $N$ notes, using a per-query seed $\sigma_q=\mathrm{md5}(q)$ for reproducibility.

\subsection{MMR (Maximal Marginal Relevance)}
MMR greedily trades off relevance against redundancy. Starting from $R\!\leftarrow\!\varnothing$, it first selects $\arg\max_i \mathrm{rel}_i$ and then repeatedly adds
\begin{equation}
\label{eq:mmr}
\begin{aligned}
i^{\star} &= \arg\max_{i\notin R}
\Big[\, \lambda\,\mathrm{rel}_i - (1-\lambda)\,\max_{j\in R} S_{ij} \,\Big], \\[2pt]
R &\leftarrow R \cup \{i^{\star}\}.
\end{aligned}
\end{equation}

until $|R|=B$. The trade-off $\lambda\in[0,1]$ interpolates between pure relevance ($\lambda{=}1$) and pure diversity ($\lambda{=}0$); we sweep $\lambda\in\{0.3,0.5,0.7,0.9\}$ and report the best.

\subsection{DPP (Determinantal Point Process)}
A $k$-DPP models \citep{kulesza2012determinantal} a probability over subsets proportional to a sub-determinant of a kernel, jointly rewarding quality and diversity. We use the quality--similarity kernel

\begin{equation}
L = \mathrm{diag}(\mathbf{q})\,S\,\mathrm{diag}(\mathbf{q}), \qquad q_i = \exp\!\big(\theta\,\mathrm{rel}_i\big),
\label{eq:dpp-kernel}
\end{equation}

where $q_i>0$ is the relevance-driven quality ($\theta$ a temperature, default $1$) and $S$ from Eq.~\eqref{eq:rel-sim} is PSD, so $L\succeq 0$. We seek the MAP configuration
\begin{equation}
R = \arg\max_{|R|=B}\ \log\det\!\big(L_{R}\big),
\label{eq:dpp-map}
\end{equation}

which is NP-hard, and approximate it with the fast greedy algorithm of \citep{chen2018fast}. Maintaining incremental Cholesky factors $\mathbf{c}_i$ and
residual gains $d_i^2$ (initialized $d_i^2=L_{ii}=q_i^2$), each step selects
\begin{equation}
j = \arg\max_{i\notin R} d_i^{2},
\qquad
R \leftarrow R \cup \{j\},
\label{eq:dpp-select}
\end{equation}
and then updates every remaining $i\notin R$ via
\begin{equation}
\label{eq:dpp-update}
\begin{aligned}
e_i &= \frac{L_{ji} - \langle \mathbf{c}_j,\mathbf{c}_i\rangle}{d_j}, \\[2pt]
\mathbf{c}_i &\leftarrow [\,\mathbf{c}_i;\, e_i\,],
\qquad
d_i^{2} \leftarrow d_i^{2} - e_i^{2}.
\end{aligned}
\end{equation}

with $d_j=\sqrt{d_j^2}$. The recursion terminates at $|R|=B$ or when no remaining note yields a positive gain ($d_i^2\!\le\!0$, i.e.\ the selected set already spans the relevant subspace).

\subsection{PPR-as-Retriever}
This baseline retrieves using the note \emph{link graph} (the VLM-induced relatedness edges) rather than the embedding affinity, testing whether graph structure alone suffices. Let $A^{\mathrm{lnk}}$ be the symmetrized binary adjacency of the note links. We form the row-stochastic transition

\begin{equation}
P_{ij} =
\begin{cases}
A^{\mathrm{lnk}}_{ij}\big/\sum_{k} A^{\mathrm{lnk}}_{ik}, & \sum_{k} A^{\mathrm{lnk}}_{ik} > 0,\\[4pt]
1/N, & \text{otherwise (dangling node)},
\end{cases}
\label{eq:ppr-transition}
\end{equation}
and a query-personalized teleport (seed) distribution from the relevances,
\begin{equation}
s_i = \frac{\exp(\mathrm{rel}_i/T)}{\sum_{k=1}^{N}\exp(\mathrm{rel}_k/T)} .
\label{eq:ppr-seed}
\end{equation}
The PPR scores are the fixed point of the power iteration

\begin{equation}
\label{eq:ppr-iter}
\begin{aligned}
\mathbf{r}^{(t+1)} &= (1-\beta)\,\mathbf{s} + \beta\,P^{\top}\mathbf{r}^{(t)}, \\[2pt]
\mathbf{r}^{(t+1)} &\leftarrow \mathbf{r}^{(t+1)}/\|\mathbf{r}^{(t+1)}\|_1,
\end{aligned}
\end{equation}

with damping $\beta{=}0.85$ and temperature $T{=}0.1$, iterated to convergence ($\|\mathbf{r}^{(t+1)}-\mathbf{r}^{(t)}\|_\infty<10^{-9}$). The evidence set is $R=\operatorname*{Top\text{-}}{B}_i\, r_i^{\star}$.

\subsection{Single-VLM (no retrieval)}
\label{app:singlevlm}
This baseline removes retrieval entirely and feeds document content directly to the reader, following the G2-Reader protocol. For query $q$ we take its first $M{=}5$ candidate documents and extract, from their MinerU parse, the set of text chunks $\mathcal{C}_q$ (recursive splitting, chunk size $3000$, overlap $300$) and the set of visual elements $\mathcal{I}_q$ (figures/tables/charts with their crops). To meet the context budget we deterministically subsample

\begin{equation}
\label{eq:svlm-sample}
\begin{aligned}
\tilde{\mathcal{C}}_q &= \mathrm{Sample}\big(\mathcal{C}_q,\ \min(K_c,|\mathcal{C}_q|);\ \sigma_q\big), \\[2pt]
\tilde{\mathcal{I}}_q &= \mathrm{Sample}\big(\mathcal{I}_q,\ \min(K_i,|\mathcal{I}_q|);\ \sigma_q\big),
\end{aligned}
\end{equation}
\begin{table*}[t]
\centering\small
\setlength{\tabcolsep}{5pt}
\caption{Hyperparameters used across all experiments. Values are the defaults used throughout; parentheses denote the ranges explored in ablations.}
\label{tab:hparams}
\begin{tabular}{ll}
\toprule
Hyperparameter & Value \\
\midrule
\multicolumn{2}{l}{\emph{Document processing \& graph}} \\
\quad Candidate documents per query & 5 (1 gold + 4 distractors) \\
\quad Text chunk size / overlap & 3000 / 300 chars \\
\quad Image local-context window & $\pm$1000 words \\
\quad Graph evolution iterations & 3 \\
\midrule
\multicolumn{2}{l}{\emph{Text encoder}} \\
\quad Model & nomic-embed-text-v1.5 \\
\quad Embedding dimension & 768 \\
\quad Normalization & $\ell_2$ \\
\midrule
\multicolumn{2}{l}{\emph{CDS retrieval}} \\
\quad Evidence budget $B$ (top-$B$) & 10 \ (\{3,5,10,15,20\}) \\
\quad Note--note affinity & $1-\cos$ (dissimilarity) \\
\quad Diversification strength $\beta$ & 1.0 \ (\{0.5,1,2,5\}) \\
\quad Query sparsity $\tau$ & 0 \ (\{0,0.3,0.5,0.6\}) \\
\quad Constraint margin $\varepsilon$ & $10^{-3}$ \\
\quad Support threshold & $10^{-6}$ \\
\quad Initialization prior & uniform \\
\quad Replicator max.\ iterations & $10{,}000$ \\
\quad Replicator tolerance $\delta$ & $10^{-7}$ \\
\midrule
\multicolumn{2}{l}{\emph{Reader (generation)}} \\
\quad Readers & Qwen3-VL-32B; Qwen2.5-VL-7B; GLM-4.1V-9B \\
\quad Decoding & greedy (temperature $0$) \\
\quad Max new tokens & 1024 \\
\quad Max text chars per node & 1500 \\
\midrule
\multicolumn{2}{l}{\emph{vLLM serving}~\citep{kwon2023efficient}} \\
\quad Tensor-parallel size & 4 (Qwen3-VL-32B) \\
\quad Max model length & 32768 (Qwen); 65536 (GLM) \\
\quad GPU memory utilization & 0.80 \\
\quad Data type & bfloat16 \\
\quad MM processor cache & disabled \\
\midrule
\multicolumn{2}{l}{\emph{Evaluation}} \\
\quad Judge model & GPT-4o-mini \\
\quad Runs per configuration & 3 (report mean\,$\pm$\,std) \\
\midrule
\multicolumn{2}{l}{\emph{Baseline selectors (ablations)}} \\
\quad MMR trade-off $\lambda$ & 0.5 \ (\{0.3,0.5,0.7,0.9\}) \\
\quad DPP quality temperature $\theta$ & 1.0 \\
\quad Prior temperature $T$ (query-softmax/diffusion) & 0.1 \ (\{0.05,0.1,0.2\}) \\
\quad PPR damping $\gamma$ & 0.85 \\
\bottomrule
\end{tabular}
\end{table*}
\begin{table*}[t]
\centering\footnotesize
\setlength{\tabcolsep}{4pt}
\caption{Initialization-prior ablation (CDS, Qwen3-VL-32B). Accuracy (\%),
mean\,$\pm$\,std over 3 runs.}
\label{tab:prior}
\resizebox{\textwidth}{!}{%
\begin{tabular}{l ccccc >{\columncolor{gray!15}}c !{\vrule width 0.8pt} c}
\toprule
Prior & SPIQA & FetaTab & PaperTab & SciGraphQA & SlideVQA & \textbf{Avg.} & MMLongBench \\
\midrule
Uniform (default)          & 78.8\,$\pm$\,0.2 & 71.0\,$\pm$\,0.6 & 65.6\,$\pm$\,0.4 & 57.4\,$\pm$\,0.4 & 62.2\,$\pm$\,0.4 & 67.0\,$\pm$\,0.2 & 45.0\,$\pm$\,0.4 \\
Query-softmax ($T{=}0.05$) & 78.4\,$\pm$\,0.1 & 70.0\,$\pm$\,0.3 & 66.0\,$\pm$\,0.0 & 58.8\,$\pm$\,0.3 & 61.6\,$\pm$\,0.8 & 67.0\,$\pm$\,0.1 & 44.8\,$\pm$\,0.2 \\
Query-softmax ($T{=}0.1$)  & 78.8\,$\pm$\,0.3 & 71.6\,$\pm$\,0.4 & 64.6\,$\pm$\,0.3 & 59.2\,$\pm$\,0.5 & 61.4\,$\pm$\,0.4 & 67.1\,$\pm$\,0.2 & 44.9\,$\pm$\,0.1 \\
Query-softmax ($T{=}0.2$)  & 78.8\,$\pm$\,0.6 & 71.7\,$\pm$\,0.3 & 65.2\,$\pm$\,0.2 & 59.1\,$\pm$\,0.8 & 61.6\,$\pm$\,0.2 & \textbf{67.3\,$\pm$\,0.3} & 45.0\,$\pm$\,0.3 \\
Graph-diffusion            & 78.0\,$\pm$\,0.4 & 71.4\,$\pm$\,0.2 & 65.6\,$\pm$\,0.6 & 59.1\,$\pm$\,0.1 & 61.6\,$\pm$\,0.1 & 67.2\,$\pm$\,0.2 & 44.4\,$\pm$\,0.3 \\
\bottomrule
\end{tabular}%
}
\end{table*}

\begin{table*}[t]
\centering\small
\setlength{\tabcolsep}{4pt}
\caption{Graph-evolution ablation (CDS, Qwen3-VL-32B). Accuracy (\%) at each
evolution iteration, computed on queries answerable in all iterations.}
\label{tab:evolution}
\begin{tabular}{lcccccc}
\toprule
Iter. & SPIQA & FetaTab & PaperTab & SciGraphQA & SlideVQA & Avg. \\
\midrule
0 (initial) & 80.2 & 70.4 & 67.0 & 57.8 & 60.3 & 67.1 \\
1           & 79.5 & 72.0 & 64.3 & 58.3 & 62.5 & 67.3 \\
2           & 77.6 & 72.3 & 64.1 & 57.0 & 63.2 & 66.9 \\
3 (final)   & 78.1 & 71.7 & 65.4 & 58.5 & 61.2 & 67.0 \\
\bottomrule
\end{tabular}
\end{table*}

with $K_c{=}K_i{=}10$ and a per-query seed $\sigma_q=\mathrm{md5}(\text{seed}\,{:}\,q_{\mathrm{id}})$ so sampling is fixed and reproducible across runs; we report mean$\pm$std over $\text{seed}\in\{42,43,44\}$. The sampled chunks are concatenated into a context $c_q=\mathrm{concat}(\tilde{\mathcal{C}}_q)$, interleaved with $\tilde{\mathcal{I}}_q$ and the question in a chain-of-thought prompt (Appendix~\ref{app:prompts}), and the answer is decoded greedily,

\begin{equation}
\hat{a} = f_{\mathrm{VLM}}\big(q,\,c_q,\,\tilde{\mathcal{I}}_q\big),
\qquad \tau = 0 .
\label{eq:svlm-answer}
\end{equation}
To avoid context overflow, the generation length is clamped to fit the server window $L_{\max}$, using a $\sim\!3$ characters/token text estimate and a fixed
$1500$-token allowance per image: 

\begin{equation}
\label{eq:svlm-clamp}
\begin{split}
T_{\mathrm{out}} = \max\!\Big(256,\ \min\!\big(8192,\
&\, L_{\max} - \tfrac{|c_q|}{3} \\
&\, - 1500\,|\tilde{\mathcal{I}}_q| - 1500 \big)\!\Big).
\end{split}
\end{equation}

\section{Hyperparameters}
\label{app:hparams}

Table~\ref{tab:hparams} lists all hyperparameters used in our experiments. Values are the defaults used throughout unless a specific ablation sweeps them (sweep ranges noted in parentheses).

\section{Additional Experiments}
\label{app:additional}
We report two further ablations on Qwen3-VL-32B: the replicator-dynamics initialization prior, and the effect of iterative graph evolution.

\paragraph{Initialization prior.} CDS solves the quadratic program in Eq.~\eqref{eq:cds_objective} via the replicator dynamics (Eq.~\eqref{eq:replicator}). 
\begin{equation}
x_i^{(t+1)}
=
x_i^{(t)}
\frac{
(M\mathbf{x}^{(t)})_i+c
}{
(\mathbf{x}^{(t)})^{\top}M\mathbf{x}^{(t)}+c
}.
\label{eq:replicator}
\end{equation}
The vector $\mathbf{x}$ can be initialized in different ways. From an initial distribution $\mathbf{x}^{(0)}$, which only determines \emph{which} local optimum the dynamics converge to, not the affinity matrix itself. We compare three choices. The first is a \emph{uniform} prior, $x_i^{(0)}=1/|\mathcal{U}|$, which is our default. The second is a \emph{query-softmax} prior that seeds each document node by its temperature-scaled relevance to the query,
\begin{table*}[t]\centering\footnotesize\setlength{\tabcolsep}{4pt}
\caption{Diversification-block ablation (CDS, Qwen3-VL-32B). The note--note affinity is set to no coupling, similarity, or dissimilarity (ours). Accuracy (\%), mean$\pm$std over 3 runs.}
\label{tab:divblock}
\resizebox{\textwidth}{!}{%
\begin{tabular}{l ccccc >{\columncolor{gray!15}}c !{\vrule width 0.8pt} c}
\toprule
note--note & SPIQA & FetaTab & PaperTab & SciGraphQA & SlideVQA & \textbf{Avg.} & MMLongBench \\
\midrule
sim ($+\cos$)                    & 75.9 $\pm$ 0.3 & 64.0 $\pm$ 0.5 & 58.9 $\pm$ 0.4 & 56.7 $\pm$ 0.3 & 59.6 $\pm$ 0.2 & 63.0 $\pm$ 0.3 & 45.4 $\pm$ 0.2 \\
\textbf{dissim ($1-\cos$, CDS)}  & \textbf{78.8 $\pm$ 0.2} & \textbf{71.0 $\pm$ 0.6} & \textbf{65.6 $\pm$ 0.4} & \textbf{57.4 $\pm$ 0.4} & \textbf{62.2 $\pm$ 0.4} & \textbf{67.0 $\pm$ 0.2} & 45.0 $\pm$ 0.4 \\
\bottomrule
\end{tabular}}
\end{table*}

\begin{table*}[t]\centering\footnotesize\setlength{\tabcolsep}{4pt}
\scriptsize
\caption{Query-row sparsification ablation (CDS, Qwen3-VL-32B). Accuracy (\%), mean$\pm$std over 3 runs.}
\label{tab:tau}
\resizebox{\textwidth}{!}{%
\begin{tabular}{l ccccc >{\columncolor{gray!15}}c !{\vrule width 0.8pt} c}
\toprule
Config & SPIQA & FetaTab & PaperTab & SciGraphQA & SlideVQA & \textbf{Avg.} & MMLongBench \\
\midrule
CDS ($\tau{=}0$)   & 78.1 $\pm$ 0.2 & 71.1 $\pm$ 0.4 & 64.5 $\pm$ 0.4 & 58.3 $\pm$ 0.4 & 61.8 $\pm$ 0.4 & 66.8 $\pm$ 0.1 & 45.1 $\pm$ 0.4 \\
CDS ($\tau{=}0.3$) & 78.5 $\pm$ 0.2 & 71.2 $\pm$ 0.8 & 65.1 $\pm$ 0.6 & 59.2 $\pm$ 0.6 & 61.9 $\pm$ 0.3 & \textbf{67.2 $\pm$ 0.1} & 44.7 $\pm$ 0.3 \\
CDS ($\tau{=}0.5$) & 77.3 $\pm$ 0.3 & 71.2 $\pm$ 0.5 & 64.0 $\pm$ 0.8 & 58.6 $\pm$ 0.1 & 62.1 $\pm$ 0.1 & 66.6 $\pm$ 0.2 & 44.4 $\pm$ 0.1 \\
CDS ($\tau{=}0.6$) & 76.6 $\pm$ 0.4 & 67.3 $\pm$ 0.2 & 65.4 $\pm$ 0.8 & 53.7 $\pm$ 0.9 & 59.6 $\pm$ 0.4 & 64.5 $\pm$ 0.1 & 43.5 $\pm$ 0.3 \\
\bottomrule
\end{tabular}}
\end{table*}

\begin{table*}[t]
\centering
\footnotesize
\setlength{\tabcolsep}{4pt}
\caption{Evidence-budget (CDS, Qwen3-VL-32B). Accuracy (\%),
mean\,$\pm$\,std over 3 runs. $B$ is the number of evidence nodes given to the reader.}
\label{tab:budget}
\begin{tabular}{lccccccc}
\toprule
$B$ & FetaTab & PaperTab & SciGraphQA & SlideVQA & SPIQA & Average & MMLongBench \\
\midrule
3  & 54.8\,$\pm$\,0.2 & 50.6\,$\pm$\,0.2 & 45.0\,$\pm$\,0.4 & 50.5\,$\pm$\,0.5 & 69.1\,$\pm$\,0.3 & 54.0\,$\pm$\,0.2 & 40.2\,$\pm$\,0.1 \\
5  & 65.7\,$\pm$\,0.5 & 57.2\,$\pm$\,0.6 & 48.3\,$\pm$\,0.6 & 58.6\,$\pm$\,0.3 & 72.4\,$\pm$\,0.3 & 60.4\,$\pm$\,0.2 & 42.1\,$\pm$\,0.2 \\
10 & 71.0\,$\pm$\,0.6 & 65.6\,$\pm$\,0.4 & 57.4\,$\pm$\,0.4 & 62.2\,$\pm$\,0.4 & 78.8\,$\pm$\,0.2 & 67.0\,$\pm$\,0.2 & 45.0\,$\pm$\,0.4 \\
15 & 74.5\,$\pm$\,0.2 & 67.6\,$\pm$\,0.5 & 61.5\,$\pm$\,0.9 & 65.3\,$\pm$\,0.3 & 79.3\,$\pm$\,0.4 & \textbf{69.6\,$\pm$\,0.1} & 46.4\,$\pm$\,0.6 \\
20 & 73.6\,$\pm$\,0.3 & 66.4\,$\pm$\,0.9 & 61.6\,$\pm$\,0.1 & 66.0\,$\pm$\,0.1 & 80.2\,$\pm$\,0.3 & \textbf{69.6\,$\pm$\,0.2} & 47.7\,$\pm$\,0.1 \\
\bottomrule
\end{tabular}
\end{table*}

\begin{table*}[t]
\centering
\setlength{\tabcolsep}{4pt}
\caption{Diversification-strength. The note--note affinity block is scaled by
\(\beta\) (\(\beta{=}1\) recovers the default CDS). Accuracy (\%),
mean\,$\pm$\,std over 3 runs; Qwen3-VL-32B reader.}
\label{tab:beta}
\resizebox{\textwidth}{!}{%
\scriptsize
\begin{tabular}{l ccccc >{\columncolor{gray!15}}c !{\vrule width 0.8pt} c}
\toprule
\(\beta\) & FetaTab & PaperTab & SciGraphQA & SlideVQA & SPIQA & \textbf{Avg.} & MMLongBench \\
\midrule
0.5 & \textbf{72.2\,$\pm$\,0.2} & 65.0\,$\pm$\,0.2 & \textbf{58.8\,$\pm$\,0.5} & \textbf{62.8\,$\pm$\,0.6} & 78.5\,$\pm$\,0.0 & \textbf{67.4\,$\pm$\,0.2} & \textbf{45.4\,$\pm$\,0.1} \\
1.0\,(CDS) & 71.0\,$\pm$\,0.6 & \textbf{65.6\,$\pm$\,0.4} & 57.4\,$\pm$\,0.4 & 62.2\,$\pm$\,0.4 & \textbf{78.8\,$\pm$\,0.2} & 67.0\,$\pm$\,0.2 & 45.0\,$\pm$\,0.4 \\
2.0 & 65.3\,$\pm$\,1.0 & 59.4\,$\pm$\,0.8 & 50.0\,$\pm$\,1.2 & 57.1\,$\pm$\,0.4 & 76.2\,$\pm$\,0.1 & 61.6\,$\pm$\,0.1 & 42.2\,$\pm$\,0.1 \\
5.0 & 42.6\,$\pm$\,0.2 & 46.1\,$\pm$\,0.3 & 14.7\,$\pm$\,0.1 & 34.0\,$\pm$\,0.1 & 51.1\,$\pm$\,0.4 & 37.7\,$\pm$\,0.1 & 35.0\,$\pm$\,0.1 \\
\bottomrule
\end{tabular}%
}
\end{table*}

\begin{equation*}
p_i = \frac{\exp(\cos(v_i,q)/T)}{\sum_j \exp(\cos(v_j,q)/T)},
\end{equation*}
which we evaluate at $T\in\{0.05,0.1,0.2\}$. The third is a \emph{graph-diffusion} prior that propagates the query-softmax seed $\mathbf{p}$ over the note link graph via Personalized PageRank,

\begin{equation*}
\mathbf{r}^{(t+1)} = (1-\gamma)\,\mathbf{p} + \gamma\,P^{\top}\mathbf{r}^{(t)},
\end{equation*}

where $P$ is the row-stochastic transition matrix of the link graph and $\gamma$ the damping factor; the converged $\mathbf{r}$ is used as $\mathbf{x}^{(0)}$. We stress that the graph-diffusion prior uses PageRank \emph{only to seed} the dynamics; the evidence is still selected by CDS. This is distinct from the PPR \emph{retriever} in Table~\ref{tab:selectors}, where PageRank \emph{replaces} CDS and notes are ranked by their PageRank scores directly. The contrast is informative: as a standalone retriever PPR is weak (54.4 average, Table~\ref{tab:selectors}), yet as an initialization it is indistinguishable from any other prior (67.2 average, Table~\ref{tab:prior}). More generally, all priors fall within $0.3$ points on the VisDoMBench average ($67.0$--$67.3$) and overlap within standard deviation on every dataset (Table~\ref{tab:prior}). CDS is thus robust to initialization---its quality is driven by the affinity, not the seed or the link-graph structure---so we adopt the parameter-free uniform prior by default.

\paragraph{Graph evolution.} The memory graph undergoes refinement over up to three evolution iterations, during which the VLM updates links and summaries. Table~\ref{tab:evolution} presents accuracy at each iteration for queries answerable in all iterations, ensuring that the observed trend reflects the effect of evolution alone. The overall impact is minimal; the VisDoMBench average changes from $67.1$ to $67.0$ across iterations, with only minor, dataset-dependent fluctuations. This suggests that evolution primarily enhances robustness rather than improving accuracy. The final graph is used in all subsequent experiments.

\subsection{Query-Row Sparsification}
\label{app:tau}
The query vertex connects to every document node weighted by non-negative cosine relevance. The threshold $\tau$ optionally prunes weak query--document edges (those with $\cos(q,v_i)<\tau$), concentrating the query anchor on its most relevant neighbours. Table~\ref{tab:tau} sweeps $\tau\in\{0,0.3,0.5,0.6\}$. Performance is flat for mild sparsification: $\tau{=}0$ and $\tau{=}0.3$ are
statistically indistinguishable on the VisDoMBench average ($66.8$ vs.\ $67.2$), with $\tau{=}0.3$ marginally ahead, indicating that removing the weakest query edges neither helps nor hurts meaningfully. Beyond this point accuracy degrades: $\tau{=}0.5$ drops to $66.6$ and $\tau{=}0.6$ to $64.5$, because an aggressive threshold starts discarding genuinely relevant evidence---the decline is sharpest on SciGraphQA ($59.2\!\rightarrow\!53.7$) and SlideVQA, whose answers draw on more, individually weaker query--node edges. Since the mild-sparsity gain is within noise and adds a hyperparameter, we use no sparsification ($\tau{=}0$) by default; the result confirms CDS is robust to this choice over a wide range.

\subsection{Sign of the Inter-Node Coupling}
\label{app:divblock}
CDS encodes diversity through the note--note block of the affinity, set to the \emph{dissimilarity} $1-\cos(v_i,v_j)$ so that similar evidence repels. To test whether this design choice---rather than merely the presence of inter-node edges---is responsible for the gains, we replace it with the opposite sign, a \emph{similarity} coupling $\max(\cos(v_i,v_j),0)$ that instead rewards selecting mutually similar notes. Table~\ref{tab:divblock} shows the sign is decisive: flipping the block to similarity drops the VisDoMBench average from $67.0$ to $63.0$, a $4.0$-point loss, and degrades every dataset (e.g., FetaTab $71.0\!\rightarrow\!64.0$, PaperTab $65.6\!\rightarrow\!58.9$). The similarity coupling drives the dominant set toward a tight cluster of near-duplicate notes, wasting the budget on redundant evidence, whereas the dissimilarity coupling spreads selection across complementary evidence while the query anchor preserves relevance. This confirms that the diversification term must \emph{repel} similar notes: it is the source of CDS's advantage, not an incidental component.

\subsection{Evidence budget:}
\label{app:evibud}
We vary the number of evidence nodes $B$ passed to the reader from $3$ to $20$ (Table~\ref{tab:budget}). Accuracy increases
monotonically with $B$: it rises steeply up to $B=10$ ($54.0 \to 67.0$ on the VisDoMBench average), gains a further $+2.6$ points at $B=15$ ($69.6$), and then plateaus on VisDoMBench at $B=20$ ($69.6$). On MMLongBench-Doc the upward trend persists slightly longer ($45.0 \to 46.4 \to 47.7$), as the single long document can absorb more evidence before saturating. The run-to-run standard deviation is small throughout ($\leq 0.9$ per dataset), so these differences are well outside noise. We use $B=10$ in all main experiments to match the evidence budget of prior work~\citep{du2026g} for a controlled comparison; the practical optimum on VisDoMBench is $B=15$. Crucially, accuracy is stable across a wide range ($B \in [10,20]$ varies by $\leq 2.6$ points), indicating that CDS is robust to the evidence-set size and does not require careful budget tuning.

\paragraph{Diversification strength.} The note--note block of our affinity, $\beta(1-\cos)$, controls how strongly the retriever favors mutually dissimilar (non-redundant) evidence relative to raw query relevance: $\beta \to 0$ removes the inter-note coupling and reduces CDS to relevance-only selection, $\beta = 1$ is the default, and larger $\beta$ amplifies the repulsion between selected notes. Table~\ref{tab:beta} sweeps this knob on the full VisDoMBench. We find a broad, flat optimum for \emph{light} diversification---$\beta = 0.5$ is marginally best (67.4 avg) and the default $\beta = 1$ is statistically indistinguishable (67.0)---indicating that a modest diversity pressure is beneficial and that CDS is insensitive to the exact value in this regime. Pushing diversification further, however, degrades accuracy sharply and monotonically ($\beta = 2 \to 61.6$, $\beta = 5 \to 37.7$), as the selected set drifts toward dissimilar but off-topic evidence (the collapse is most severe on SciGraphQA, $58.8 \to 14.7$). In other words, diversity helps only up to the point where it begins to override relevance.

\section{Prompts}
\label{app:prompts}
We list every prompt used in our pipeline and evaluation. Placeholders in braces (e.g.\ \texttt{\{question\}}) are filled at run time.
\onecolumn 
\begin{promptbox}{Prompt: Text Element Analysis (memory-note construction)}
\begin{promptverb}
Generate a structured analysis of the following content by:
1. Identifying the most salient keywords (focus on nouns, verbs, and key concepts)
2. Extracting core themes, concepts and arguments
3. Creating relevant categorical tags

Format the response as a JSON object:
{
  "keywords": [  // specific, distinct keywords, ordered most->least important;
                 // at least three, avoid redundancy
  ],
  "summary":     // one sentence: main topic/domain + key points; concise
  ,
  "tags": [      // broad categories/themes (domain, format, type); >=3, non-redundant
  ]
}

Content for analysis: {content}
\end{promptverb}
\end{promptbox}

\begin{promptbox}{Prompt: Visual Element Analysis (figures/tables, from MinerU crops)}
\begin{promptverb}
Generate a structured analysis of the visual elements in the provided image.
You are also given the image's surrounding text context (before/after) and its caption.

Instructions:
1) Use the context ONLY to aid understanding of the image's role; do not quote or
   rely on it unless it aligns with what is visible or stated in the caption. Base
   the summary primarily on the visual evidence and the caption.
2) Keywords MUST include exact in-image terms: labels, legends, axis titles,
   category names, and domain-specific terms; preserve their exact wording.
3) If the caption has an index (e.g., "Figure 1", "Table 2"), begin the summary by
   formalizing it ("Figure X -- ...", "Table Y -- ...") then describe concisely.

Format the response as a JSON object:
{
  "keywords": [ // exact in-image labels/legends/axis titles/terms; >=3, non-redundant ],
  "summary":    // start with "Figure X -- ..."/"Table Y -- ..." if indexed;
                // describe the visual elements specifically
  ,
  "tags": [ // broad domain/format/type categories; >=3, non-redundant ]
}

(Two few-shot examples are provided: a labeled map and an ablation table.)

This is context: {context} and caption: {caption}
\end{promptverb}
\end{promptbox}

\begin{promptbox}{Prompt: Memory Evolution (graph link induction + note refinement)}
\begin{promptverb}
You are an AI memory-evolution agent managing a knowledge base. Given a memory
note (content, summary, keywords) and its neighboring notes, decide its evolution.

The memory note content: {content}
Summary: {context}
Keywords: {keywords}
The {neighbor_number} neighboring notes:
{neighbors}

Determine:
1. Which neighboring notes should be linked to this note?
2. Should this note's summary/keywords be updated given those relationships?
3. If so, what are the new summary and keywords?

Connect two notes ONLY for a specific logical relationship: direct reference,
causal, part-whole, conceptual elaboration, temporal sequence,
contrastive/comparative, hierarchical, or contextual dependency. DO NOT connect
notes that merely share keywords, domain, surface similarity, document adjacency,
or generic concepts. Order connections by relevance (most relevant first).

When updating: make the note more precise and distinctive; describe ONLY this
note's content (no comparative language about other notes); prefer specific,
distinctive keywords; keep the summary self-contained and <= 30 words. If the note
already captures its unique content, do not update.

Return JSON:
{
  "suggested_connections": ["neighbor_memory_ids"],
  "should_update": true or false,
  "new_summary": "new summary",
  "new_keywords": ["keyword_1", ..., "keyword_n"]
}
\end{promptverb}
\end{promptbox}

\begin{promptbox}{Prompt: Answer Generation (reader VLM, CDS pipeline)}
\begin{promptverb}
[System]
You are answering a question using the provided document context. Each context
item comes from a single document. Some items are text passages; others are
figures or tables sent as images (each accompanied by a short caption/summary).

Rules:
 1. Use ONLY the provided context. Do not invent facts.
 2. Quote table numbers, figure numbers, named entities, and numeric values exactly.
 3. If the answer is not in the context, reply exactly:
    "Not found in the provided context."
 4. Answer concisely. No preamble.

[User]
Question:
{question}

Context:
--- Note 1  (id=.., weight=.., type=text|image) ---
{text passage  OR  image caption/summary + <image>}
--- Note 2  ... ---
...
Answer:
\end{promptverb}
\end{promptbox}

\begin{promptbox}{Prompt: Single-VLM Baseline (no retrieval, chain-of-thought)}
\begin{promptverb}
Please read the following text and the attached images and answer the question below.

<text>
{context}
</text>

What is the correct answer to this question: {question}

Format your response as follows:
"<reason>detailed reason for your answer here</reason>
 <answer>the correct answer here</answer>".
Make sure your answer is comprehensive and covers all important information related
to the question. If directly relevant information is not provided, give your best
answer based on the available context (without mentioning that you lack information).
\end{promptverb}
\end{promptbox}

\begin{promptbox}{Prompt: Answer Evaluation (LLM-as-judge, GPT-4o-mini)}
\begin{promptverb}
You are an expert evaluator assessing answers from a RAG system.
Task: judge whether the generated answer correctly responds to the question,
given the expected answer.

Question:        {question}
Expected Answer: {gold_answers}
Generated Answer:{assistant_answer}

Accuracy (0 or 1): 1 if factually correct and aligned with the expected answer; 0 otherwise.

Apply uniformly:
- Judge factual correctness, not style/formatting. Ignore capitalization,
  punctuation, whitespace, quotes, articles, markup, abbreviation vs full form,
  synonyms, singular/plural.
- Partial-match: correct if the answer contains ALL key facts of the expected
  answer and adds only non-contradictory context.
- Concise-match: a terser label/phrase/number that conveys the same key assertion
  is correct (unless the gold has multiple distinct claims the prediction omits).
- Abstractive-elaboration: paraphrase preserving subject-predicate-object is correct.
- Role-preservation: if entities are kept but the relation direction is reversed,
  mark incorrect.
- Numeric: equivalent up to unit/format (0.82 == 82%); signed deltas must match.
- List: correct if all key items present (any order).
- Multi-gold: correct if it matches ANY one acceptable answer.
- Entity: must refer to the same subject as the question/expected answer.

"Not answerable": correct only if BOTH say not answerable; otherwise incorrect.

Output (JSON only):
{
  "accuracy": 0 or 1,
  "reasoning": "brief explanation"
}
\end{promptverb}
\end{promptbox}

\end{document}